\documentclass{svjour3}

\title{A Mechanised Proof of G\"odel's Incompleteness Theorems using Nominal Isabelle}
\author{Lawrence C. Paulson}
\institute{Computer Laboratory, University of Cambridge, England\\ \email{lp15@cam.ac.uk}}
\date{}

\usepackage{amsmath,amssymb,bm,url,graphicx}
\usepackage{isabelle,isabellesym}
 \newtheorem{thm}{Theorem}

\newcommand\Pf{\mathop{\rm Pf\,}}
\newcommand\supp{\mathop{\rm supp}}
\newcommand\quot[1]{\ulcorner#1\urcorner}
\newcommand\tuple[1]{\langle#1\rangle}

\newcommand\ex[2]{\exists#1\,[#2]}
\newcommand\all[2]{\forall#1\,[#2]}
\let\bimp=\leftrightarrow
\let\ts=\thinspace
\let\eats=\lhd

\begin{document}
\def\makeheadbox{} 
\maketitle

\begin{abstract}
An Isabelle/HOL formalisation of G\"odel's two incompleteness theorems 
is presented. The work follows \'Swierczkowski's detailed proof of the theorems
using hereditarily finite (HF) set theory \cite{swierczkowski-finite}. Avoiding the usual arithmetical encodings of syntax eliminates the necessity to formalise elementary number theory within an embedded logical calculus. The Isabelle formalisation uses two separate treatments of variable binding: the nominal package \cite{urban-general} is shown to scale to a development of this complexity, while de Bruijn indices~\cite{debruijn72} turn out to be ideal for coding syntax. Critical details of the Isabelle proof are described, in particular gaps and errors found in the literature.
\end{abstract}

\section{Introduction}

This paper describes mechanised proofs of G\"odel's incompleteness theorems \cite{goedel-formally},
including the first mechanised proof of the second incompleteness theorem. Very informally, these results can be stated as follows:

\begin{thm}[First Incompleteness Theorem]
If $L$ is a consistent theory capable of formalising a sufficient amount of elementary mathematics,
then there is a sentence $\delta$ such that neither $\delta$ nor $\neg\delta$ is a theorem of~$L$, and moreover, $\delta$ is true.%
\footnote{Meaning, $\delta$ (which has no free variables) is true in the standard model for~$L$.}
\end{thm}

\begin{thm}[Second Incompleteness Theorem]
If $L$ is as above and $\mathop{\rm Con}(L)$ is a sentence stating that $L$ is consistent,
then $\mathop{\rm Con}(L)$ is not a theorem of~$L$.
\end{thm}

Both of these will be presented formally below. Let us start to examine what these theorems actually assert.
They concern a consistent formal system, say~$L$, based on first-order logic with
some additional axioms: G\"odel chose Peano arithmetic (PA) \cite{goedel32b},
but hereditarily finite (HF) set theory is an alternative \cite{swierczkowski-finite}, used here.
The first theorem states that any such axiomatic system must be
incomplete, in the sense that some sentence can neither be proved nor disproved. The expedient
of adding that sentence as an axiom merely creates a new axiomatic system, for which there is
another undecidable sentence. The theorem can be strengthened to allow infinitely many additional axioms,
provided there is an effective procedure to recognise whether a given formula is an axiom or not.

The second incompleteness theorem asserts that the consistency of~$L$ cannot be proved in~$L$ itself. Even to state this theorem rigorously requires first defining the concept of provability in~$L$; the necessary series of definitions amounts to a computer program that occupies many pages. Although the same definitions are used to prove the first incompleteness theorem, they are at least not needed to state that theorem. The original rationale for this project was a logician's suggestion that the second incompleteness theorem had never been proved rigorously. Having completed this project, I sympathise with his view; most published proofs contain substantial gaps and use cryptic notation.

Both incompleteness theorems are widely misinterpreted, both in popular culture
and even by some mathematicians. The first incompleteness theorem is often taken to imply
that mathematics cannot be formalised, when evidently it has been,
this paper being one of numerous instances.
It has also been used to assert that human intelligence can perceive truths
(in particular, the truth of~$\delta$, the undecidable sentence) that no computer will ever understand.
Franz{\'e}n \cite{franzen-guide} surveys and demolishes many of these fallacies.
The second incompleteness theorem destroyed Hilbert's hope that the consistency of quite strong theories
might be proved even in Peano arithmetic. It also tells us, for example, that the axioms of
set theory do not imply the existence of an inaccessible cardinal, as that would yield a model
for set theory itself.

The first incompleteness theorem has been proved with machine assistance at least three times before. The first time (surprisingly early: 1986) was by Shankar \cite{shankar-phd,shankar94}, using Nqthm. Then in 2004, O'Connor \cite{oconnor-incompleteness} (using Coq) 
and Harrison (using HOL Light)%
\footnote{Proof files at \url{http://code.google.com/p/hol-light/}, directory \texttt{Arithmetic}}
each proved versions of the theorem.
The present proof,
conducted using Isabelle/HOL, is novel in adopting nominal syntax \cite{urban-general}
for formalising variable binding in the syntax of~$L$, while using de Bruijn notation \cite{debruijn72} for coding those formulas. Despite using two different treatments of variable binding, the necessary representation theorem for formulas is not difficult to prove.
It is not clear that other treatments of higher-order abstract syntax could make this claim.
These proofs can be seen as an extended demonstration
of the power of nominal syntax, while at the same time vindicating de Bruijn indexing in some situations.

The machine proofs are fairly concise at under 12,400 lines for both theorems.%
\footnote{This is approximately as long as Isabelle's theory of Kurzweil-Henstock gauge integration.}
The paper presents highlights of the proof,
commenting on the advantages and disadvantages of the nominal framework and HF set theory. An overview of the project from a logician's perspective has appeared elsewhere \cite{paulson-incompl-logic}. The proof
reported here closely follows a detailed exposition by \'Swierczkowski \cite{swierczkowski-finite}.
His careful and detailed proofs were indispensable, despite some errors and omissions, which are reported below.
For the first time, we have complete, formal proofs of both theorems. They take the form
of structured Isar proof scripts \cite{Incompleteness-AFP} that can be examined interactively.

The remainder of the paper presents background material (Sect.\ts\ref{sec:background}) before outlining the development itself: the proof calculus (Sect.\ts\ref{sec:theorems}), the coding of the calculus within itself (Sect.\ts\ref{sec:coding}) and finally the first theorem (Sect.\ts\ref{sec:G-I}). Technical material relating to the second theorem are developed (Sect.\ts\ref{sec:pseudo}) then the theorem is presented and discussed (Sect.\ts\ref{sec:G-II}). Finally, the paper concludes (Sect.\ts\ref{sec:conclusions}).

\section{Background} \label{sec:background}

Isabelle/HOL \cite{isa-tutorial} is an interactive theorem prover for higher-order logic.
This formalism can be seen as extending a polymorphic typed first-order logic with a functional language
in which recursive datatypes and functions can be defined. Extensive documentation
is available.\footnote{\url{http://isabelle.in.tum.de/documentation.html}}

For interpreting the theorem statements presented below, it is important to note that a theorem
that concludes $\psi$ from the premises $\phi_1$, \ldots, $\phi_n$ may be expressed in a variety of
equivalent forms. The following denote precisely the same theorem:
\begin{quote}
\isa{\isasymlbrakk\(\phi_1\);\ \ldots;\ \(\phi_n\)\isasymrbrakk\ \isasymLongrightarrow\ \(\psi\)}

\isa{\(\phi_1\)\ \isasymLongrightarrow\ \(\cdots\)\ \isasymLongrightarrow\ \(\phi_n\)\ \isasymLongrightarrow\ \(\psi\)}

\isa{\isakeyword{assumes}\ \(\phi_1\)\ \isakeyword{and}\ \ldots\ \isakeyword{and}\ \(\phi_n\)\ \isakeyword{shows}\ \(\psi\)}
\end{quote}
If the conclusion of a theorem involves the keyword \isakeyword{obtains}, then there is an implicit
existential quantification. The following two theorems are logically equivalent.
\begin{quote}
\isa{\(\phi\)\ \isasymLongrightarrow\ \(\exists x.\)\ \(\psi_1\land\ldots\land\psi_n\)}

\isa{\isakeyword{assumes}\ \(\phi\)\ \isakeyword{obtains}\ \(x\)\ \isakeyword{where}\ \(\psi_1\) \ldots\ \(\psi_n\)}
\end{quote}

Other background material for this paper includes an outline of G\"odel's proof,
an introduction to hereditarily finite set theory and some examples of Nominal Isabelle.

\subsection{G\"odel's Proof}

Much of G\"odel's proof may be known to many readers, but it will be useful
to list the milestones here, for reference.
\begin{enumerate}
\item
A first-order \textit{deductive calculus} is formalised, including the syntax of terms and formulas, substitution, and semantics (evaluation). The calculus includes axioms to define Peano arithmetic or some alternative,
such as the HF set theory used here.
There are inference rules for propositional and quantifier reasoning.
We write $H\vdash A$ to mean that $A$ can be proved from $H$ (a set of formulas)
in the calculus.

\item
Meta-theory is developed to relate \textit{truth and provability}. The need for tedious proof constructions
in the deductive calculus is minimised through a meta-theorem stating that a class of true formulas
are theorems of that calculus. One way to do this is through the notion of $\Sigma$ formulas,
which are built up from atomic formulas using $\lor$, $\land$, $\exists$ and bounded universal
quantification. Then the key result is
\begin{equation}
 \text{If $\alpha$ is a true $\Sigma$ sentence, then\ } {}\vdash \alpha. 
\label{eqn:truth}
\end{equation}

\item
A \textit{system of coding} is set up within the formalised first-order theory. The code of a formula $\alpha$
is written $\quot {\alpha}$ and is a term of the calculus itself.

\item
In order to \textit{formalise the calculus within itself}, predicates to recognise codes are defined, including terms and formulas, and syntactic operations on them.
Next, predicates are defined to recognise individual axioms and inference rules, then a sequence of such logical steps.
We obtain a predicate $\Pf$,
where ${}\vdash \Pf {\quot {\alpha}}$ expresses that the formula $\alpha$ has a proof.
The key result is 
\begin{equation}
{}\vdash \alpha \iff {}\vdash \Pf {\quot {\alpha}}. \label{eqn:pfc}
\end{equation}
All of these developments must be completed before the second incompleteness theorem can even be stated.
\item
G\"odel's \emph{first} incompleteness theorem is obtained by constructing a formula $\delta$ that is
provably equivalent (within the calculus) to the formal statement that $\delta$ is not provable.
It follows (provided the calculus is consistent) that neither $\delta$ nor its negation can be proved, although $\delta$ is true in the semantics.

\item
G\"odel's \emph{second} incompleteness theorem requires the following crucial lemma:
\[ \text{If $\alpha$ is a $\Sigma$ sentence, then\ } {}\vdash \alpha\to \Pf {\quot {\alpha}}. \]
This is an internalisation of theorem~(\ref{eqn:truth}) above.
It is proved by induction over the construction of $\alpha$ as a $\Sigma$ formula.
This requires generalising the statement above to allow the formula $\alpha$ to contain free variables.
The technical details are complicated, and lengthy deductions in the calculus seem to be essential.
\end{enumerate}
The proof sketched above incorporates numerous improvements over G\"odel's original version.
G\"odel proved only the left-to-right direction of the equivalence~(\ref{eqn:pfc}) and required a stronger assumption than consistency, namely $\omega$-consistency.

\subsection{Hereditarily Finite Set Theory} \label{sec:hf-set-theory}

G\"odel first proved his incompleteness theorems in a first-order theory of Peano arithmetic \cite{goedel32b}.
O'Connor and Harrison do the same, while Shankar and I have both chosen a formalisation
of the hereditarily finite (HF) sets. Although each theory can be formally interpreted in the other,
meaning that they are of equivalent strength, the HF theory is more convenient, as it can express
all set-theoretic constructions that do not require infinite sets. An HF set is
a finite set of HF sets, and this recursive definition can be captured by the following three axioms:
\begin{gather}
z=0 \bimp \all{x}{x\not\in z} \tag{HF1}\\
z=x\lhd y \bimp \all{u}{u\in z\bimp u\in x\lor u=y} \tag{HF2}\\
\phi(0) \land \all{xy}{\phi(x)\land\phi(y)\to\phi(x\lhd y)}\to \all{x}{\phi(x)} \tag{HF3}
\end{gather}
The first axiom states that 0 denotes the empty set.
The second axiom defines the binary operation symbol~$\lhd$ (pronounced ``eats'') 
to denote insertion into a set, so that $x\lhd y = x\cup \{y\}$.
The third axiom is an induction scheme, and states that every set is created by a finite number of applications of
0 and~$\lhd$. 

The machine proofs of the incompleteness theorems rest on an Isabelle theory of the hereditarily finite sets. To illustrate the syntax, here are the three basic axioms as formalised in Isabelle. The type of such sets is called \isa{hf}, and is constructed such that the axioms
above can be proved.
\begin{isabelle}
\isacommand{lemma}\ hempty\_iff:\ "z=0\ \isasymlongleftrightarrow \ (\isasymforall x.\ \isasymnot \ x\ \isactrlbold \isasymin \ z)"\isanewline
\isacommand{lemma}\ hinsert\_iff:\ "z\ =\ x\ \isasymtriangleleft \ y\ \isasymlongleftrightarrow \ (\isasymforall u.\ u\ \isactrlbold \isasymin \ z\ \isasymlongleftrightarrow \ u\ \isactrlbold \isasymin \ x\ |\ u\ =\ y)"\isanewline
\isacommand{lemma}\ hf\_induct\_ax:\ "\isasymlbrakk P\ 0;\ \isasymforall x.\ P\ x\ \isasymlongrightarrow \ (\isasymforall y.\ P\ y\ \isasymlongrightarrow \ P\ (x\ \isasymtriangleleft \ y))\isasymrbrakk \ \isasymLongrightarrow \ P\ x"
\end{isabelle}
The same three axioms, formalised within Isabelle as a deep embedding, form the basis for the incompleteness proofs. Type \isa{hf} and its associated operators serve as the standard model for the embedded HF set theory.

HF set theory is equivalent to Zermelo-Frankel (ZF) set theory with the axiom of infinity negated. Many of the Isabelle definitions and theorems were taken, with minor modifications,
from Isabelle/ZF \cite{paulson-set-I}. Familiar concepts such as union, intersection, set difference,
power set, replacement, ordered pairing and foundation can be derived from the axioms shown above
\cite{swierczkowski-finite}. A few of these derivations need to be repeated---with infinitely greater effort---in the internal calculus.

Ordinals in HF are simply natural numbers, where $n=\{0,1,\ldots,n-1\}$.
Their typical properties (for example, that they are linearly ordered) have the same proofs
as in ZF set theory. \'Swierczkowski's proofs \cite{swierczkowski-finite} are sometimes
more elegant, 
and addition on ordinals is obtained through a general notion of addition of sets \cite{kirby-addition}.
Finally, there are about 400 lines of material concerned with relations, functions and finite sequences. This is needed
to reason about the coding of syntax for the incompleteness theorem.

\subsection{Isabelle's Nominal Package} \label{sec:nominal}

For the incompleteness theorems, we are concerned with formalising the syntax of
first-order logic. Variable binding is a particular issue: it is well known that technicalities
relating to bound variables and substitution have caused errors in published proofs and complicate
formal treatments. O'Connor \cite{oconnor-phd} reports severe difficulties in his proofs.

The \emph{nominal} approach \cite{gabbay-pitts,pitts-nominal} to formalising variable binding 
(and other sophisticated uses of variable names) is based on a theory of permutations over names. A primitive concept is
\emph{support}: $\supp(x)$ has a rather technical definition involving permutations, but is equivalent
in typical situations to the set of free names in~$x$. We also write $a\mathbin{\sharp} x$ for
$a\not\in \supp(x)$, saying ``$a$ is fresh for $x$''. Isabelle's nominal package \cite{urban-nominal,urban-general} supports these concepts
through commands such as \isacommand{nominal\_datatype} to introduce types, 
\isacommand{nominal\_primrec} to declare primitive recursive functions and \isacommand{nominal\_induct} to perform structural induction. Syntactic constructions involving variable binding are identified up to $\alpha$-conversion, using a quotient construction. These mechanisms generally succeed at emulating informal standard conventions for variable names. In particular, we can usually assume that the bound variables we encounter never clash with other variables.

The best way to illustrate these ideas is by examples. The following datatype defines
the syntax of terms in the HF theory:
\begin{isabelle}
\isacommand{nominal\_datatype}\ tm\ =\ Zero\ |\ Var\ name\ |\ Eats\ tm\ tm
\end{isabelle}
The type \isa{name} (of variable names) has been created using the nominal framework. The members of this type constitute a countable set of uninterpreted atoms. The function \isa{nat\_of\_name} is a bijection between this type and the type of natural numbers.

Here is the syntax of HF formulas, which are $t\in u$, $t=u$, $\phi\lor\psi$, $\neg\phi$
or $\ex{x}{\phi}$:
\begin{isabelle}
\isacommand{nominal\_datatype}\ fm\ =\isanewline
\ \ \ \ Mem\ tm\ tm\ \ \ \ (\isakeyword{infixr}\ "IN"\ 150)\isanewline
\ \ |\ Eq\ tm\ tm\ \ \ \ \ (\isakeyword{infixr}\ "EQ"\ 150)\isanewline
\ \ |\ Disj\ fm\ fm\ \ \ (\isakeyword{infixr}\ "OR"\ 130)\isanewline
\ \ |\ Neg\ fm\isanewline
\ \ |\ Ex\ x::name\ f::fm\ \isakeyword{binds}\ x\ \isakeyword{in}\ f
\end{isabelle}
In most respects, this \emph{nominal} datatype behaves exactly like a standard algebraic datatype.
However, the bound variable name designated by \isa{x} above is not significant: no function can be defined to return the name of a variable bound using \isa{Ex}.

Substitution of a term \isa{x} for a variable \isa{i} is defined as follows:
\begin{isabelle}
\isacommand{nominal\_primrec}\ subst\ ::\ "name\ \isasymRightarrow \ tm\ \isasymRightarrow \ tm\
\isasymRightarrow \ tm"\isanewline
\ \ \isakeyword{where}\isanewline
\ \ \ "subst\ i\ x\ Zero\ \ \ \ \ \ \ =\ Zero"\isanewline
\ |\ "subst\ i\ x\ (Var\ k)\ \ \ \ =\ (if\ i=k\ then\ x\ else\ Var\ k)"\isanewline
\ |\ "subst\ i\ x\ (Eats\ t\ u)\ =\ Eats\ (subst\ i\ x\ t)\ (subst\ i\ x\ u)"
\end{isabelle}
Unfortunately, most recursive definitions involving \isacommand{nominal\_primrec} must be followed by
a series of proof steps, verifying that the function uses names legitimately.
Occasionally, these proofs (omitted here) require subtle reasoning involving nominal primitives.

Substituting the term \isa{x} for the variable \isa{i} in the formula \isa{A}
is written \isa{A(i::=x)}.
\begin{isabelle}
\isacommand{nominal\_primrec}\ \ subst\_fm\ ::\ "fm\ \isasymRightarrow \ name\ \isasymRightarrow \ tm\
\isasymRightarrow \ fm"\isanewline
\ \ \isakeyword{where}\isanewline
\ \ \ \ Mem:\ \ "(Mem\ t\ u)(i::=x)\ \ =\ Mem\ (subst\ i\ x\ t)\ (subst\ i\ x\ u)"\isanewline
\ \ |\ Eq:\ \ \ "(Eq\ t\ u)(i::=x)\ \ \ =\ Eq\ \ (subst\ i\ x\ t)\ (subst\ i\ x\ u)"\isanewline
\ \ |\ Disj:\ "(Disj\ A\ B)(i::=x)\ =\ Disj\ (A(i::=x))\ (B(i::=x))"\isanewline
\ \ |\ Neg:\ \ "(Neg\ A)(i::=x)\ \ \ \ =\ Neg\ (A(i::=x))"\isanewline
\ \ |\ Ex:\ \ \ "atom\ j\ \isasymsharp \ (i,\ x)\ \isasymLongrightarrow \ (Ex\ j\ A)(i::=x)\ =\ Ex\ j\
(A(i::=x))"
\end{isabelle}
Note that the first seven cases (considering the two substitution functions collectively) are straightforward
structural recursion. In the final case, we see a precondition, \isa{atom\ j\ \isasymsharp \ (i,\ x)},
to ensure that the substitution is legitimate within the formula \hbox{\isa{Ex\ j\ A}}. There is no way to define the contrary case, where the bound variable clashes. We would have to eliminate any such clash, renaming the bound variable by applying an appropriate permutation to the formula. Thanks to the nominal framework, such explicit renaming steps are rare. 

This style of formalisation is more elegant than traditional textbook definitions that do include the variable-clashing case. It is much more elegant than including renaming of the bound variable as part of the definition itself. Such ``definitions'' are really implementations, and greatly complicate proofs.

The commutativity of substitution (two distinct variables, each fresh for the opposite term) is easily proved.
\begin{isabelle}
\isacommand{lemma}\ subst\_fm\_commute2\ [simp]:\isanewline
\ \ "\isasymlbrakk atom\ j\ \isasymsharp \ t; atom\ i\ \isasymsharp \ u;
i\ \isasymnoteq \ j\isasymrbrakk\ \isasymLongrightarrow \ (A(i::=t))(j::=u)\ =\ (A(j::=u))(i::=t)"\isanewline
\ \ \isacommand{by}\ (nominal\_induct\ A\ avoiding:\ i\ j\ t\ u\ rule:\ fm.strong\_induct)\ auto
\end{isabelle}
The proof is by \emph{nominal induction} on the formula~\isa{A}: the proof method for structural induction over a nominal datatype. Compared with ordinary induction, \isa{nominal\_induct} takes account of the freshness of bound variable names. The phrase \isa{avoiding:\ i\ j\ t\ u}
is the formal equivalent of the convention that when we consider the case of the existential formula
\isa{Ex\ k\ A}, the bound variable \isa{k} can be assumed to be fresh with respect to the terms mentioned.
This convention is formalised by four additional assumptions \isa{atom\ k\ \isasymsharp \ i}, \isa{atom\ k\ \isasymsharp \ j}, etc.; they ensure that substitution will be well-defined over this existential quantifier, making the proof easy. 

This and many similar facts have two-step proofs, \isa{nominal\_induct} followed by \isa{auto}. In contrast, O'Connor needed to combine three substitution lemmas
(including the one above) in a giant mutual induction involving 1,900 lines of Coq.
He blames the renaming step in substitution and suggests that a form of simultaneous
substitution might have avoided these difficulties \cite[\S4.3]{oconnor-phd}. An alternative, using traditional bound variable names, is to treat substitution 
not as a function but as a relation that holds only when no renaming is necessary. Bound variable renaming is then an independent operation. I briefly tried this idea, which allowed reasonably straightforward proofs of substitution properties, but ultimately nominal looked like a better option.

\section{Theorems, $\Sigma$ Formulas, Provability} \label{sec:theorems}

The first milestone in the proof of the incompleteness theorems is the development
of a first-order logical calculus equipped with enough meta-theory to guarantee that
some true formulas are theorems. The previous section has already presented the definitions
of the terms and formulas of this calculus. The terms are for HF set theory, and the formulas
are defined via a minimal set of connectives from which others can be defined. 

\subsection{A Sequent Calculus for HF Set Theory}

Compared with a textbook presentation of G\"odel's proof, a machine development must include an effective proof system, one that can actually be used to prove non-trivial theorems.

\subsubsection{Semantics}

The semantics of terms and formulas are given by the 
obvious recursive function definitions, which yield sets and Booleans, respectively.
These functions accept an environment mapping variables to values. The null environment
maps all variables to 0, and is written~\isa{e0}. It involves the types \isa{finfun} (for finite functions)
\cite{lochbihler-finfuns} and \isa{hf} (for HF sets).
\begin{isabelle}
\isacommand{definition}\ e0\ ::\ "(name,\ hf)\ finfun"\ \ \ \ %
\isamarkupcmt{the null environment}
\isanewline
\ \ \isakeyword{where}\ "e0\ \isasymequiv \ finfun\_const\ 0"
\end{isabelle}

\begin{isabelle}
\isacommand{nominal\_primrec}\ eval\_tm\ ::\ "(name,\ hf)\ finfun\ \isasymRightarrow \ tm\ \isasymRightarrow \ hf"\isanewline
\ \ \isakeyword{where}\isanewline
\ \ \ "eval\_tm\ e\ Zero\ =\ 0"\isanewline
\ |\ "eval\_tm\ e\ (Var\ k)\ =\ finfun\_apply\ e\ k"\isanewline
\ |\ "eval\_tm\ e\ (Eats\ t\ u)\ =\ eval\_tm\ e\ t\ \isasymtriangleleft \ eval\_tm\ e\ u"
\end{isabelle}

There are two things to note in the semantics of formulas.
First, the special syntax \isa{\isasymlbrakk t\isasymrbrakk e} abbreviates \isa{eval\_tm\ e\ t}.
Second, in the semantics of the quantifier \isa{Ex}, note how the formula \isa{atom\ k\ \isasymsharp \ e}
asserts that the bound variable, \isa{k}, is not currently given a value by the environment, \isa{e}.
\begin{isabelle}
\isacommand{nominal\_primrec}\ eval\_fm\ ::\ "(name,\ hf)\ finfun\ \isasymRightarrow \ fm\ \isasymRightarrow \ bool"\isanewline
\ \ \isakeyword{where}\isanewline
\ \ \ "eval\_fm\ e\ (t\ IN\ u)\ \isasymlongleftrightarrow \ \isasymlbrakk t\isasymrbrakk e\ \isasymin\ \isasymlbrakk u\isasymrbrakk e"\isanewline
\ |\ "eval\_fm\ e\ (t\ EQ\ u)\ \isasymlongleftrightarrow \ \isasymlbrakk t\isasymrbrakk e\ =\ \isasymlbrakk u\isasymrbrakk e"\isanewline
\ |\ "eval\_fm\ e\ (A\ OR\ B)\ \isasymlongleftrightarrow \ eval\_fm\ e\ A\ \isasymor \ eval\_fm\ e\ B"\isanewline
\ |\ "eval\_fm\ e\ (Neg\ A)\ \isasymlongleftrightarrow \ (\isachartilde \ eval\_fm\ e\ A)"\isanewline
\ |\ "atom\ k\ \isasymsharp \ e\ \isasymLongrightarrow \ eval\_fm\ e\ (Ex\ k\ A)\ \isasymlongleftrightarrow \ (\isasymexists x.\ eval\_fm\ (finfun\_update\ e\ k\ x)\ A)"
\end{isabelle}
This yields the Tarski truth definition for the standard model of HF set theory. In particular, \isa{eval\_fm\ e0\ A} denotes the truth of the sentence~\isa{A}.

\subsubsection{Axioms}

\'Swierczkowski \cite{swierczkowski-finite} specifies a standard Hilbert-style calculus, with two rules of inference and several axioms or axiom schemes. The latter include \textit{sentential axioms}, defining the behaviour of disjunction and negation:
\begin{isabelle}
\isacommand{inductive\_set}\ boolean\_axioms\ ::\ "fm\ set"\isanewline
\ \ \isakeyword{where}\isanewline
\ \ \ \ Ident:\ \ \ \ \ "A\ IMP\ A\ \isasymin \ boolean\_axioms"\isanewline
\ \ |\ DisjI1:\ \ \ \ "A\ IMP\ (A\ OR\ B)\ \isasymin \ boolean\_axioms"\isanewline
\ \ |\ DisjCont:\ \ "(A\ OR\ A)\ IMP\ A\ \isasymin \ boolean\_axioms"\isanewline
\ \ |\ DisjAssoc:\ "(A\ OR\ (B\ OR\ C))\ IMP\ ((A\ OR\ B)\ OR\ C)\ \isasymin \ boolean\_axioms"\isanewline
\ \ |\ DisjConj:\ \ "(C\ OR\ A)\ IMP\ (((Neg\ C)\ OR\ B)\ IMP\ (A\ OR\ B))\ \isasymin \ boolean\_axioms"
\end{isabelle}
Here \'Swierczkowski makes a tiny error, expressing the last axiom scheme as
\[ (\phi\lor\psi)\land (\neg\phi\lor\mu) \to\psi\lor\mu. \]
Because $\land$ is defined in terms of $\lor$, while this axiom helps to define~$\lor$, this formulation is unlikely to work. The Isabelle version eliminates $\land$ in favour of nested implication.

There are four primitive \textit{equality axioms}, shown below in mathematical notation. They express reflexivity as well as substitutivity for equality, membership and the eats operator. They are not schemes but single formulas containing specific free variables. Creating an instance of an axiom for specific terms 
(which might involve the same variables)
requires many renaming steps to insert fresh variables, before substituting for them one term at a time.
\begin{align*}
x_1 &= x_1 \\
(x_1=x_2)\land (x_3=x_4) &\to [(x_1 = x_3)\to (x_2 = x_4)] \\
(x_1=x_2)\land (x_3=x_4) &\to [(x_1\in x_3)\to (x_2\in x_4)] \\
(x_1=x_2)\land (x_3=x_4) &\to [x_1\lhd x_3 = x_2 \lhd x_4] 
\end{align*}

There is also a \textit{specialisation} axiom scheme, of the form $\phi(t/x)\to\exists x\,\phi$:
\begin{isabelle}
\isacommand{inductive\_set}\ special\_axioms\ ::\ "fm\ set"\ \isakeyword{where}\isanewline
\ \ I:\ "A(i::=x)\ IMP\ (Ex\ i\ A)\ \isasymin \ special\_axioms"
\end{isabelle}

There are the axioms HF1 and HF2 for the set theory, while HF3 (\textit{induction}) is formalised as an axiom scheme: 
\begin{isabelle}
\isacommand{inductive\_set}\ induction\_axioms\ ::\ "fm\ set"\ \isakeyword{where}\isanewline
\ \ ind:\isanewline
\ \ "atom\ (j::name)\ \isasymsharp \ (i,A)\isanewline
\ \ \ \isasymLongrightarrow \ A(i::=Zero)\isanewline
\ \ \ \ \ \ \ IMP\ ((All\ i\ (All\ j\ (A\ IMP\ (A(i::=\ Var\ j)\ IMP\ A(i::=\ Eats(Var\ i)(Var\ j))))))\isanewline
\ \ \ \ \ \ IMP\ (All\ i\ A))\isanewline
\ \ \ \ \isasymin \ induction\_axioms"
\end{isabelle}

Axiom schemes are conveniently introduced using \isacommand{inductive\_set}, simply to express set comprehensions, even though there is no actual induction.
 
\subsubsection{Inference System}

The axiom schemes shown above, along with inference rules for modus ponens and existential instantiation,%
\footnote{From $A\to B$ infer $\exists x A\to B$, for $x$ not free in $B$.}
are combined to form the following inductive definition of theorems:
\begin{isabelle}
\isacommand{inductive}\ hfthm\ ::\ "fm\ set\ \isasymRightarrow \ fm\ \isasymRightarrow \ bool"\ (\isakeyword{infixl}\ "\isasymturnstile "\ 55)\isanewline
\ \ \isakeyword{where}\isanewline
\ \ \ \ Hyp:\ \ \ \ "A\ \isasymin \ H\ \isasymLongrightarrow \ H\ \isasymturnstile \ A"\isanewline
\ \ |\ Extra:\ \ "H\ \isasymturnstile \ extra\_axiom"\isanewline
\ \ |\ Bool:\ \ \ "A\ \isasymin \ boolean\_axioms\ \isasymLongrightarrow \ H\ \isasymturnstile \ A"\isanewline
\ \ |\ Eq:\ \ \ \ \ "A\ \isasymin \ equality\_axioms\ \isasymLongrightarrow \ H\ \isasymturnstile \ A"\isanewline
\ \ |\ Spec:\ \ \ "A\ \isasymin \ special\_axioms\ \isasymLongrightarrow \ H\ \isasymturnstile \ A"\isanewline
\ \ |\ HF:\ \ \ \ \ "A\ \isasymin \ HF\_axioms\ \isasymLongrightarrow \ H\ \isasymturnstile \ A"\isanewline
\ \ |\ Ind:\ \ \ \ "A\ \isasymin \ induction\_axioms\ \isasymLongrightarrow \ H\ \isasymturnstile \ A"\isanewline
\ \ |\ MP:\ \ \ \ \ "H\ \isasymturnstile \ A\ IMP\ B\ \isasymLongrightarrow \ H'\ \isasymturnstile \ A\ \isasymLongrightarrow \ H\ \isasymunion \ H'\ \isasymturnstile \ B"\isanewline
\ \ |\ Exists:\ "H\ \isasymturnstile \ A\ IMP\ B\ \isasymLongrightarrow \ atom\ i\isasymsharp B\ \isasymLongrightarrow \ \isasymforall C\isasymin H.\ atom\ i\isasymsharp C\ \isasymLongrightarrow \ H\ \isasymturnstile \ (Ex\ i\ A)\ IMP\ B"
\end{isabelle}
A minor deviation from \'Swierczkowski is \isa{extra\_axiom}, 
which is abstractly specified to be an arbitrary true formula. This means that the proofs will be conducted with respect to an arbitrary finite extension of the HF theory. The first major deviation from \'Swierczkowski is the introduction of rule~\isa{Hyp}, with a set of assumptions. It would be virtually impossible to prove anything in his Hilbert-style proof system, and it was clear from the outset that lengthy proofs within the calculus might be necessary. Introducing \isa{Hyp} generalises the notion of provability, allowing the development of a sort of sequent calculus, in which long but tolerably natural proofs can be constructed.

It is worth mentioning that \'Swierczkowski's definitions and proofs fit together very tightly, deviations often being a cause for later regret. One example, concerning an inference rule for substitution, is mentioned at the end of Sect.\ts\ref{sec:deduction}. Another example is that some tricks that simplify the proof of the first incompleteness theorem turn out to complicate the proof of the second.

The soundness of the calculus above is trivial to prove by induction. The deduction theorem
is also straightforward, the only non-trivial case being the one for the \isa{Exists}
inference rule.
The induction formula is stated as follows:
\begin{isabelle}
\isacommand{lemma}\ deduction\_Diff:\ \isakeyword{assumes}\ "H\ \isasymturnstile \ B"\ \isakeyword{shows}\ "H\ -\ \isacharbraceleft C\isacharbraceright \ \isasymturnstile \ C\ IMP\ B"
\end{isabelle}
This directly yields the standard formulation of the deduction theorem:
\begin{isabelle}
\isacommand{theorem}\ deduction:\ \isakeyword{assumes}\ "insert\ A\ H\ \isasymturnstile \ B"\ \isakeyword{shows}\ "H\ \isasymturnstile \ A\ IMP\ B"
\end{isabelle}
And this is a sequent rule for implication.

Setting up a usable sequent calculus requires much work. 
The corresponding Isabelle theory file, which starts with the definitions of terms and formulas and ends with a sequent formulation of the HF induction rule, is nearly 1,600 lines long. Deriving natural sequent calculus rules from the sentential and equality axioms  requires lengthy chains of steps. Even in the final derived sequent calculus, equalities can only be applied one step at a time. 

For another example of difficulty, consider the following definition:
\begin{isabelle}
\isacommand{definition}\ Fls\ \ \isakeyword{where}\ "Fls\ \isasymequiv \ Zero\ IN\ Zero"
\end{isabelle}
Proving that \isa{Fls} has the properties of falsehood is surprisingly tricky. The relevant
axiom, HF1, is formulated using universal quantifiers, which are defined as negated existentials;
deriving the expected properties of universal quantification seems to require something like
\isa{Fls} itself. 

The derived sequent calculus has specialised rules to operate on conjunctions, disjunctions, etc., in the hypothesis part of a sequent. They are crude, but good enough. Used with Isabelle's automatic tactics, they ease somewhat the task of constructing formal HF proofs. 
Users can extend Isabelle with proof procedures coded in ML, and better automation for the calculus might thereby be achieved.
At the time, such a side-project did not seem to be worth the effort.

\subsection{A Formal Theory of Functions} \label{sec:functions}

Recursion is not available in HF set theory, and recursive functions must be constructed explicitly.
Each recursive computation is expressed in terms of the existence of a sequence $(s_i)_{i\le k}$
such that $s_i$ is related to $s_m$ and $s_n$ for $m$, $n<i$. 
Moreover, a sequence is formally a relation rather than a function.
In the metalanguage, we write \isa{app s k} for $s_k$, 
governed by the theorem
\begin{isabelle}
\isacommand{lemma}\ app\_equality:\ "hfunction\ s\ \isasymLongrightarrow \ \isasymlangle x,y\isasymrangle \ \isasymin\ s\ \isasymLongrightarrow \ app\ s\ x\ =\ y"
\end{isabelle}

The following two functions express the recursive definition of sequences, as needed for the
G\"odel development:
\begin{isabelle}
\ \ "Builds\ B\ C\ s\ l\ \isasymequiv \ B(app\ s\ l)\ \isasymor \ (\isasymexists m\isasymin l.\ \isasymexists n\isasymin l.\ C(app\ s\ l)\ (app\ s\ m)\ (app\ s\ n))"\vskip1ex
\ \ "BuildSeq\ B\ C\ s\ k\ y\ \isasymequiv \ LstSeq\ s\ k\ y\ \isasymand \ (\isasymforall l\isasymin succ\ k.\ Builds\ B\ C\ s\ l)"
\end{isabelle}
The statement \isa{Builds\ B\ C\ s\ l} constrains element \isa{l} of sequence \isa{s}, namely
\isa{app\ s\ l}. We have either \isa{B(app\ s\ l)}, or \isa{C\ (app\ s\ l)\ (app\ s\ m)\ (app\ s\ n))} where \isa{m\isasymin l} and \isa{n\isasymin l}. For the natural numbers, set membership coincides with
the less-than relation. Therefore, we are referring to a sequence $s$ and element $s_l$
where either the base case $B(s_l)$ holds, or else the recursive step $C(s_l,s_m,s_n)$ for $m$, $n<l$.
The statement \isa{BuildSeq\ B\ C\ s\ k\ y} states that the sequence \isa{s} has been constructed in this 
way right up to the value \isa{app s k}, or in other words, $s_k$, where $y=s_k$.

To formalise the basis for this approach requires a series of definitions in the HF calculus, introducing the subset relation, ordinals (which are simply natural numbers), ordered pairs, relations with a given domain, etc.
Foundation (the well-foundedness of the membership relation) must also be proved,
which in turn requires additional definitions. A few highlights are shown below.

The \emph{subset relation} is defined, with infix syntax \isa{SUBS}, with the help of
\isa{All2}, the bounded universal quantifier. 
\begin{isabelle}
\isacommand{nominal\_primrec}\ Subset\ ::\ "tm\ \isasymRightarrow \ tm\ \isasymRightarrow \ fm"\ \ \ (\isakeyword{infixr}\ "SUBS"\ 150)\isanewline
\ \ \isakeyword{where}\ "atom\ z\ \isasymsharp \ (t,\ u)\ \isasymLongrightarrow \ t\ SUBS\ u\ =\ All2\ z\ t\ ((Var\ z)\ IN\ u)"
\end{isabelle}
In standard notation, this says $t\subseteq u = (\forall z\in t) [z\in u]$.
The definition uses \isacommand{nominal\_primrec}, even though it is not recursive, because it requires \isa{z} to be fresh with respect to the terms~\isa{t} and~\isa{u}, among other nominal-related technicalities.

\emph{Extensionality} is taken as an axiom in traditional set theories, but in HF it can be proved
by induction. However, many straightforward properties of the subset relation
must first be derived.
\begin{isabelle}
\isacommand{lemma}\ Extensionality:\ "H\ \isasymturnstile \ x\ EQ\ y\ IFF\ x\ SUBS\ y\ AND\ y\ SUBS\ x"
\end{isabelle}

\emph{Ordinals} will be familiar to set theorists. The definition is the usual one, and shown below mainly as an example of a slightly more complicated HF formula. Two variables, \isa{y} and \isa{z}, must be fresh for each other and \isa{x}.
\begin{isabelle}
\isacommand{nominal\_primrec}\ OrdP\ ::\ "tm\ \isasymRightarrow \ fm"\isanewline
\ \ \isakeyword{where}\ "\isasymlbrakk atom\ y\ \isasymsharp \ (x,\ z);\ atom\ z\ \isasymsharp \ x\isasymrbrakk \ \isasymLongrightarrow \isanewline
\ \ \ \ OrdP\ x\ =\ All2\ y\ x\ ((Var\ y)\ SUBS\ x\ \ AND\isanewline
\ \ \ \ \ \ \ \ \ \ \ \ \ All2\ z\ (Var\ y)\ ((Var\ z)\ SUBS\ (Var\ y)))"
\end{isabelle}

The formal definition of a \emph{function} (as a single-valued set of pairs) is subject to several complications. As we shall see in Sect.\ts\ref{sec:sigma} below,
all definitions must use $\Sigma$ formulas, which requires certain non-standard
formulations. In particular, $x\not=y$ is not a $\Sigma$ formula in general,
but it can be expressed as $x<y\lor y<x$ if $x$ and $y$ are ordinals.
The following primitive is used extensively when coding the syntax of HF within itself.
\begin{isabelle}
\isacommand{nominal\_primrec}\ LstSeqP\ ::\ "tm\ \isasymRightarrow \ tm\ \isasymRightarrow \ tm\ \isasymRightarrow \ fm"\isanewline
\ \ \isakeyword{where}\isanewline
\ \ \ "LstSeqP\ s\ k\ y\ =\ OrdP\ k\ AND\ HDomain\_Incl\ s\ (SUCC\ k)\ AND\isanewline
\ \ \ \ \ \ \ \ \ \ \ \ \ \ \ \ \ \ \ \ HFun\_Sigma\ s\ AND\ HPair\ k\ y\ IN\ s"
\end{isabelle}
Informally, \isa{LstSeqP}~$s$~$k$~$y$ means that $s$ is a non-empty sequence
whose domain includes the set $\{0,\ldots,k\}$ (which is the ordinal $k+1$: the sequence is at least that long). 
Moreover, $y=s_k$;
that would be written \isa{\isasymlangle k,y\isasymrangle \ \isasymin\ s} in the metalanguage,
but becomes \isa{HPair\ k\ y\ IN\ s} in the HF calculus, 
as seen above.

\'Swierczkowski \cite{swierczkowski-finite} prefers slightly different definitions,
specifying the domain to be exactly~$k$, where $k>0$ and $y=s_{k-1}$. 
The definition shown above simplifies the proof
of the first incompleteness theorem, but complicates the proof of the second, in particular
because they allow a sequence to be longer than necessary.

This part of the development consists mainly of proofs in the HF calculus, and is nearly 1,300 lines long.

\subsection{$\Sigma$ Formulas and Provability} \label{sec:sigma}

G\"odel had the foresight to recognise the value of minimising the need to write explicit formal proofs,
without relying on the assumption that certain proofs could ``obviously'' be formalised. Instead, he developed enough meta-theory to prove that these proofs existed.
One approach for this \cite{boolos-provability,swierczkowski-finite}
relies on the concept of $\Sigma$ formulas. These are inductively defined
to include all formulas of the form $t\in u$, $t=u$, $\alpha\lor\beta$, $\alpha\land\beta$, 
$\exists x\,\alpha$ and $(\forall x\in t)\,\alpha$.
(These are closely related to the $\Sigma_1$ formulas of the arithmetical hierarchy.)
It follows by induction on this construction that every true $\Sigma$ sentence has a formal proof. 
Intuitively, the reasoning is that
the atomic cases can be calculated, the Boolean cases can be done recursively,
and the bounded universal quantifier can be replaced by a finite conjunction.
The existential case holds because the semantics of $\exists x\,\alpha$ yields a specific
witnessing value, again allowing an appeal to the induction hypothesis.

The $\Sigma$ formula approach is a good fit to the sort of formulas used in the coding of syntax.
In these formulas, universal quantifiers have simple upper bounds, typically a variable
giving the length of a sequence, while existential variables are unbounded.
G\"odel's original proofs required all quantifiers to be bounded. Existential quantifiers were bounded by complicated expressions requiring deep and difficult arithmetic justifications.
Boolos presents similar material in a more modern form \cite[p.\ts41]{boolos-provability}.
Relying exclusively on $\Sigma$ formulas avoids these complications, but instead some straightforward properties have to be proven formally in the HF calculus.

A complication is that proving the second incompleteness theorem requires
 another induction over $\Sigma$ formulas. 
To minimise that proof effort, it helps to use as restrictive a definition as possible. The \emph{strict} $\Sigma$ formulas
consist of all formulas of the form $x\in y$, $\alpha\lor\beta$, $\alpha\land\beta$, 
$\exists x\,\alpha$ and $(\forall x\in y)\,\alpha$. Here, $x$ and~$y$ are not general terms, but variables. We further stipulate $y$ not free in $\alpha$ in $(\forall x\in y)\,\alpha$; then in the main induction leading to the second incompleteness theorem, Case~2 of Lemma~9.7 \cite{swierczkowski-finite} can be omitted.
\begin{isabelle}
\isacommand{inductive}\ ss\_fm\ ::\ "fm\ \isasymRightarrow \ bool"\ \isakeyword{where}\isanewline
\ \ \ \ MemI:\ \ "ss\_fm\ (Var\ i\ IN\ Var\ j)"\isanewline
\ \ |\ DisjI:\ "ss\_fm\ A\ \isasymLongrightarrow \ ss\_fm\ B\ \isasymLongrightarrow \ ss\_fm\ (A\ OR\ B)"\isanewline
\ \ |\ ConjI:\ "ss\_fm\ A\ \isasymLongrightarrow \ ss\_fm\ B\ \isasymLongrightarrow \ ss\_fm\ (A\ AND\ B)"\isanewline
\ \ |\ ExI:\ \ \ "ss\_fm\ A\ \isasymLongrightarrow \ ss\_fm\ (Ex\ i\ A)"\isanewline
\ \ |\ All2I:\ "ss\_fm\ A\ \isasymLongrightarrow \ atom\ j\ \isasymsharp \ (i,A)\ \isasymLongrightarrow \ ss\_fm\ (All2\ i\ (Var\ j)\ A)"
\end{isabelle}
Now, a $\Sigma$ formula is by definition one that is provably equivalent (in HF)
to some strict $\Sigma$ formula containing no additional free variables.
In another minor oversight, \'Swierczkowski  omits the free variable condition, but it is necessary.
\begin{isabelle}
\isacommand{definition}\ Sigma\_fm\ ::\ "fm\ \isasymRightarrow \ bool"\isanewline
\ \ \isakeyword{where}\ "Sigma\_fm\ A\ \isasymlongleftrightarrow \ (\isasymexists B.\ ss\_fm\ B\ \isasymand\ supp\ B\ \isasymsubseteq \ supp\ A\ \isasymand\ \isacharbraceleft \isacharbraceright \ \isasymturnstile \ A\ IFF\ B)"
\end{isabelle}

As always, \'Swierczkowski's exposition is valuable, but far from complete. Showing that $\Sigma$ formulas include $t\in u$, $t=u$ and $(\forall x\in t)\,\alpha$ for all terms $t$ and~$u$ (and not only for variables) is far from obvious. These necessary facts are not even stated clearly. A crucial insight is to focus on proving that $t\in u$ and $t\subseteq u$ are $\Sigma$ formulas. Consideration of the cases  $t\in 0$,  $t\in u_1\eats u_2$, $0\subseteq u$, $t_1\eats t_2\subseteq u$ shows that each reduces to false, true or a combination of simpler uses of $\in$ or~$\subseteq$. This observation suggests proving that $t\in u$ and $t\subseteq u$ are $\Sigma$ formulas by mutual induction on the combined sizes of $t$ and~$u$. 
\begin{isabelle}
\isacommand{lemma}\ Subset\_Mem\_sf\_lemma:\isanewline
\ \ "size\ t\ +\ size\ u\ <\ n\ \isasymLongrightarrow \ Sigma\_fm\ (t\ SUBS\ u)\ \isasymand \ Sigma\_fm\ (t\ IN\ u)"
\end{isabelle}
The identical argument turns out to be needed for the second incompleteness theorem itself, formalised this time within the HF calculus. This coincidence should not be that surprising, as it is known that the second theorem could in principle be shown by formalising the first theorem within its own calculus.

Now that we have taken care of $t\subseteq u$, proving that $t=u$ is a $\Sigma$ formula is trivial by extensionality, and the one remaining objective is $(\forall x\in t)\,\alpha$. But with equality available, we can reduce this case to the strict $\Sigma$ formula $(\forall x\in y)\,\alpha$ with the help of a lemma:
\begin{isabelle}
\isacommand{lemma}\ All2\_term\_Iff:\ "atom\ i\ \isasymsharp \ t\ \isasymLongrightarrow \ atom\ j\ \isasymsharp \ (i,t,A)\ \isasymLongrightarrow \ \isanewline
\ \ \ \ \ \ \ \ \ \ \ \ \ \ \ \ \ \ \isacharbraceleft \isacharbraceright \ \isasymturnstile \ (All2\ i\ t\ A)\ IFF\ Ex\ j\ (Var\ j\ EQ\ t\ AND\ All2\ i\ (Var\ j)\ A)"
\end{isabelle}
This is simply $(\forall x\in t)\,A \bimp \ex{y}{y=t \land (\forall x\in y)\,A}$ expressed in the HF calculus, where it is easily proved. We could prove that $(\forall x\in t)\,\alpha$ is a $\Sigma$ formula by induction on~$t$, but this approach leads to complications.

Virtually all predicates defined for the G\"odel development are carefully designed to take the form of $\Sigma$ formulas. 
Here are two examples; most such lemmas hold immediately by the construction of the given formula.
\begin{isabelle}
\isacommand{lemma}\ Subset\_sf:\ "Sigma\_fm\ (t\ SUBS\ u)"\vskip1ex
\isacommand{lemma}\ LstSeqP\_sf:\ "Sigma\_fm\ (LstSeqP\ t\ u\ v)"
\end{isabelle}

The next milestone asserts that if $\alpha$ is a ground $\Sigma$ formula (and therefore a sentence) and $\alpha$ evaluates to true, then $\alpha$ is a theorem. The proof is by induction on the size of the formula, and then by case analysis on its outer form. The case $t\in u$ falls to a mutual induction with $t\subseteq u$ resembling the one shown above. The case $(\forall x\in t)\,\alpha$ is effectively expanded to a conjunction.
\begin{isabelle}
\isacommand{theorem}\ Sigma\_fm\_imp\_thm:\ "\isasymlbrakk Sigma\_fm\ \isasymalpha;\ ground\_fm\ \isasymalpha;\ eval\_fm\ e0\ \isasymalpha\isasymrbrakk \ \isasymLongrightarrow \ \isacharbraceleft \isacharbraceright \ \isasymturnstile \ \isasymalpha"
\end{isabelle}
Every true $\Sigma$ sentence is a theorem. This crucial meta-theoretic result is used eight times in the development. Without it, gigantic explicit HF proofs would be necessary.

\section{Coding Provability in HF Within Itself} \label{sec:coding}

The key insight leading to the proof of G\"odel's theorem is that a sufficiently strong logical calculus
can represent its syntax within itself, and in particular, the property of a given formula being provable.
This task divides into three parts: coding the syntax, defining predicates to describe the coding
and finally, defining predicates to describe the inference system.

\subsection{Coding Terms, Formulas, Abstraction and Substitution}

In advocating the use of HF over PA, {\'S}wierczkowski begins by emphasising the ease of coding syntax:
\begin{quote}
It is at hand to code the variables $x_1$, $x_2$, $\ldots$ simply by the ordinals 1, 2, $\ldots$.
The constant 0 can be coded as 0, and the remaining 6 symbols as $n$-tuples of 0s, 
say $\in$ as $\tuple{0,0}$, etc. And here ends the arbitrariness of coding, which is so unpleasant when languages are arithmetized. \cite[p.\ts5]{swierczkowski-finite}
\end{quote}
The adequacy of these definitions is easy to prove in HF itself.
The full list is as follows:
$\quot{0}=0$, $\quot{x_i}=i+1$, $\quot{\in}=\tuple{0,0}$, $\quot{\lhd}=\tuple{0,0,0}$,
$\quot{=}=\tuple{0,0,0,0}$,
$\quot{\lor}=\tuple{0,0,0,0,0}$,
$\quot{\neg}=\tuple{0,0,0,0,0,0}$,
$\quot{\exists}=\tuple{0,0,0,0,0,0,0}$.
We have a few differences from {\'S}wierczkowski: $\quot{x_i}=i+1$ because our variables start at zero, 
and for the $k$th de Bruijn index we use
$\tuple{\tuple{0,0,0,0,0,0,0,0},k}$.
Obviously $\in$ means nothing by itself, so $\quot{\in}=\tuple{0,0}$
really means $\quot{t\in u}=\tuple{\tuple{0,0},\quot t,\quot u}$, etc. Note that nests of $n$-tuples terminated by ordinals can be decomposed uniquely.

De Bruijn equivalents of terms and formulas are then declared. To repeat:
de Bruijn syntax is used for coding, for which it is ideal, allowing the simplest possible
definitions of abstraction and substitution. Although it destroys readability, encodings are never readable anyway. Using nominal here is out of the question. The entire theory of nominal Isabelle would need to be formalised within the embedded calculus. Quite apart from the work involved, the necessary equivalence classes would be infinite sets, which are not available in HF\@.

The strongest argument for HF is that the mathematical basis of its coding scheme is simply ordered pairs defined in the standard set-theoretic way. An elementary formal argument justifies this. In contrast, the usual arithmetic encoding relies on either the Chinese remainder theorem or unique prime factorisation. This fragment of number theory would have to be formalised within the embedded calculus in order to reason about encoded formulas, which is necessary to prove the second incompleteness theorem. It must be emphasised that proving anything in the calculus (where such luxuries as a simplifier, recursion and even function symbols are not available) is much more difficult than proving the same result in a proof assistant.

\subsubsection{Introducing de Bruijn Terms and Formulas}

De Bruijn terms resemble the type \isa{tm} declared in Sect.\ts\ref{sec:nominal}, but include the \isa{DBInd} constructor for bound variable indices
as well as the \isa{DBVar} constructor for free variables.
\begin{isabelle}
\isacommand{nominal\_datatype}\ dbtm\ =\ DBZero\ |\ DBVar\ name\ |\ DBInd\ nat\ |\ DBEats\ dbtm\ dbtm
\end{isabelle}
De Bruijn formula contructors involve no explicit variable binding, creating an apparent similarity between \isa{DBNeg} and \isa{DBEx}, although the latter creates an implicit variable binding scope.
\begin{isabelle}
\isacommand{nominal\_datatype}\ dbfm\ =\isanewline
\ \ \ \ DBMem\ dbtm\ dbtm\isanewline
\ \ |\ DBEq\ dbtm\ dbtm\isanewline
\ \ |\ DBDisj\ dbfm\ dbfm\isanewline
\ \ |\ DBNeg\ dbfm\isanewline
\ \ |\ DBEx\ dbfm
\end{isabelle}
How this works should become clear as we consider how terms and formulas are translated into their de Bruijn equivalents. To begin with, we need a lookup function taking a list of names (representing variables bound in the current context, innermost first) and a name to be looked up. The integer~\isa{n}, initially 0, is the index to substitute if the name is next in the list.
\begin{isabelle}
\isacommand{fun}\ lookup\ ::\ "name\ list\ \isasymRightarrow \ nat\ \isasymRightarrow \ name\ \isasymRightarrow \ dbtm"\isanewline
\ \ \isakeyword{where}\isanewline
\ \ \ \ "lookup\ []\ n\ x\ =\ DBVar\ x"\isanewline
\ \ |\ "lookup\ (y\ \#\ ys)\ n\ x\ =\ (if\ x\ =\ y\ then\ DBInd\ n\ else\ (lookup\ ys\ (Suc\ n)\ x))"
\end{isabelle}

To translate a term into de Bruijn format, the key step is to resolve name references using \isa{lookup}. Names bound in the local environment are replaced by the corresponding indices, while other names are left as they were.
\begin{isabelle}
\isacommand{nominal\_primrec}\ trans\_tm\ ::\ "name\ list\ \isasymRightarrow \ tm\ \isasymRightarrow \ dbtm"\isanewline
\ \ \isakeyword{where}\isanewline
\ \ \ "trans\_tm\ e\ Zero\ =\ DBZero"\isanewline
\ |\ "trans\_tm\ e\ (Var\ k)\ =\ lookup\ e\ 0\ k"\isanewline
\ |\ "trans\_tm\ e\ (Eats\ t\ u)\ =\ DBEats\ (trans\_tm\ e\ t)\ (trans\_tm\ e\ u)"
\end{isabelle}

Noteworthy is the final case of \isa{trans\_fm}, which requires the bound variable~\isa{k}
in the quantified formula \isa{Ex k A} to be fresh with respect to~\isa{e}, our list of previously-encountered bound variables. In the recursive call, \isa{k} is added to the list, which therefore consists of distinct names.
\begin{isabelle}
\isacommand{nominal\_primrec}\ trans\_fm\ ::\ "name\ list\ \isasymRightarrow \ fm\ \isasymRightarrow \ dbfm"\isanewline
\ \ \isakeyword{where}\isanewline
\ \ \ "trans\_fm\ e\ (Mem\ t\ u)\ =\ DBMem\ (trans\_tm\ e\ t)\ (trans\_tm\ e\ u)"\isanewline
\ |\ "trans\_fm\ e\ (Eq\ t\ u)\ \ =\ DBEq\ (trans\_tm\ e\ t)\ (trans\_tm\ e\ u)"\isanewline
\ |\ "trans\_fm\ e\ (Disj\ A\ B)\ =\ DBDisj\ (trans\_fm\ e\ A)\ (trans\_fm\ e\ B)"\isanewline
\ |\ "trans\_fm\ e\ (Neg\ A)\ \ \ =\ DBNeg\ (trans\_fm\ e\ A)"\isanewline
\ |\ "atom\ k\ \isasymsharp \ e\ \isasymLongrightarrow \ trans\_fm\ e\ (Ex\ k\ A)\ =\ DBEx\ (trans\_fm\ (k\#e)\ A)"
\end{isabelle}

Syntactic operations for de Bruijn notation tend to be straightforward, as there are no bound variable names that might clash. Comparisons with previous formalisations of the $\lambda$-calculus may be illuminating, but the usual lifting operation \cite{Nipkow-CR,norrish-deBruijn} is unnecessary. That is because the HF calculus does not allow reductions anywhere, as in the $\lambda$-calculus. Substitutions only happen at the top level and never within deeper bound variable contexts. For us, substitution is the usual operation of replacing a variable by a term, which contains no bound variables. (Substitution can alternatively be defined to replace a de Bruijn index by a term.)

The special de Bruijn operation is \textit{abstraction}. This replaces every occurrence of a given free variable in a term or formula by a de Bruijn index, in preparation for binding. For example, abstracting the formula \isa{DBMem (DBVar x) (DBVar y)} over the variable \isa{y} yields \isa{DBMem (DBVar x) (DBInd 0)}. This is actually ill-formed, but attaching a quantifier yields the de Bruijn formula
\begin{isabelle}
DBEx (DBMem (DBVar x) (DBInd 0)),
\end{isabelle}
representing $\ex {y} {x\in y}$.
Abstracting this over the free variable \isa{x} and attaching another quantifier yields 
\begin{isabelle}
DBEx (DBEx (DBMem (DBInd 1) (DBInd 0))),
\end{isabelle}
which is the formula $\ex {x y} {x\in y}$. An index of~1 has been substituted in order to skip over the inner binder.

\subsubsection{Well-Formed de Bruijn Terms and Formulas}

With the de Bruijn approach, an index of 0 designates the innermost enclosing binder, while an index of 1 designates the next-innermost binder, etc. (Here, the only binder is \isa{DBEx}.) If every index has a matching binder  (the index $i$ must be nested within at least $i+1$ binders), then the term or formula is \textit{well-formed}. Recall the examples of abstraction above, where a binder must be attached afterwards.

In particular, as our terms do not contain any binding constructs, a \textit{well-formed term} may contain no de Bruijn indices. In contrast to more traditional notions of logical syntax, if you take a well-formed formula and view one of its subformulas or subterms in isolation, it will not necessarily be well-formed. The situation is analogous to extracting a fragment of a program, removing it from necessary enclosing declarations.

The property of being a well-formed de Bruijn term or formula is defined inductively. The syntactic predicates defined below recognise such well-formed formulas. A well-formed de Bruijn term has no indices (\isa{DBInd}) at all:
\begin{isabelle}
\isacommand{inductive}\ wf\_dbtm\ ::\ "dbtm\ \isasymRightarrow \ bool"\isanewline
\ \ \isakeyword{where}\isanewline
\ \ \ \ Zero:\ \ "wf\_dbtm\ DBZero"\isanewline
\ \ |\ Var:\ \ \ "wf\_dbtm\ (DBVar\ name)"\isanewline
\ \ |\ Eats:\ \ "wf\_dbtm\ t1\ \isasymLongrightarrow \ wf\_dbtm\ t2\ \isasymLongrightarrow \ wf\_dbtm\ (DBEats\ t1\ t2)"
\end{isabelle}

A trivial induction shows that for every well-founded de Bruijn term there is an equivalent standard term. The only cases to be considered (as per the definition above) are \isa{Zero}, \isa{Var} and \isa{Eats}.
\begin{isabelle}
\isacommand{lemma}\ wf\_dbtm\_imp\_is\_tm:\isanewline
\ \ \isakeyword{assumes}\ "wf\_dbtm\ x"\isanewline
\ \ \isakeyword{shows}\ "\isasymexists t::tm.\ x\ =\ trans\_tm\ []\ t"
\end{isabelle}

A well-formed de Bruijn formula is constructed from other well-formed terms and formulas,
and indices can only be introduced by applying abstraction (\isa{abst\_dbfm}) over a given \isa{name}
to another well-formed formula, in the existential case. Specifically, the \isa{Ex} clause below states that, starting with a well-formed formula~\isa{A}, abstracting over some \isa{name} and applying \isa{DBEx} to the result yields another well-formed formula.
\begin{isabelle}
\isacommand{inductive}\ wf\_dbfm\ ::\ "dbfm\ \isasymRightarrow \ bool"\isanewline
\ \ \isakeyword{where}\isanewline
\ \ \ \ Mem:\ \ \ "wf\_dbtm\ t1\ \isasymLongrightarrow \ wf\_dbtm\ t2\ \isasymLongrightarrow \ wf\_dbfm\ (DBMem\ t1\ t2)"\isanewline
\ \ |\ Eq:\ \ \ \ "wf\_dbtm\ t1\ \isasymLongrightarrow \ wf\_dbtm\ t2\ \isasymLongrightarrow \ wf\_dbfm\ (DBEq\ t1\ t2)"\isanewline
\ \ |\ Disj:\ \ "wf\_dbfm\ A1\ \isasymLongrightarrow \ wf\_dbfm\ A2\ \isasymLongrightarrow \ wf\_dbfm\ (DBDisj\ A1\ A2)"\isanewline
\ \ |\ Neg:\ \ \ "wf\_dbfm\ A\ \isasymLongrightarrow \ wf\_dbfm\ (DBNeg\ A)"\isanewline
\ \ |\ Ex:\ \ \ \ "wf\_dbfm\ A\ \isasymLongrightarrow \ wf\_dbfm\ (DBEx\ (abst\_dbfm\ name\ 0\ A))"
\end{isabelle}
This definition formalises the allowed forms of construction, rather than stating explicitly that every index must have a matching binder.

A refinement must be mentioned. \textit{Strong} nominal induction (already seen above, Sect.\ts\ref{sec:nominal}) formalises the assumption that bound variables revealed by induction can be assumed not to clash with other variables. This is set up automatically for nominal datatypes, but here requires a manual step. The command \isacommand{nominal\_inductive} sets up strong induction for \isa{name} in the \isa{Ex} case of the inductive definition above; we must prove that \isa{name} is not significant according to the nominal theory, and then get to assume that \isa{name} will not clash. This step (details omitted) seems to be necessary in order to complete some inductive proofs about \isa{wf\_dbfm}.

\subsubsection{Quoting Terms and Formulas}

It is essential to remember that G\"odel encodings are terms (having type~\isa{tm}), not sets or numbers. 
Textbook presentations identify
terms with their denotations for the sake of clarity, but this can be confusing.
The undecidable formula contains an encoding of itself in the form of a term. 
First, we must define codes for de Bruijn terms and formulas.
\begin{isabelle}
\isacommand{function}\ quot\_dbtm\ ::\ "dbtm\ \isasymRightarrow \ tm"\isanewline
\ \ \isakeyword{where}\isanewline
\ \ \ "quot\_dbtm\ DBZero\ =\ Zero"\isanewline
\ |\ "quot\_dbtm\ (DBVar\ name)\ =\ ORD\_OF\ (Suc\ (nat\_of\_name\ name))"\isanewline
\ |\ "quot\_dbtm\ (DBInd\ k)\ \ \ \ =\ HPair\ (HTuple\ 6)\ (ORD\_OF\ k)"\isanewline
\ |\ "quot\_dbtm\ (DBEats\ t\ u)\ =\ HPair\ (HTuple\ 1)\ (HPair\ (quot\_dbtm\ t)\ (quot\_dbtm\ u))"
\end{isabelle}

The codes of real terms and formulas (for which we set up the overloaded
\isa{\isasymlceil \(\cdots\)\isasymrceil} syntax) are obtained by first translating them to their de Bruijn equivalents and then encoding. We finally obtain facts such as the following:
\begin{isabelle}
\isacommand{lemma}\ quot\_Zero:\ "\isasymlceil Zero\isasymrceil \ =\ Zero"\isanewline
\isacommand{lemma}\ quot\_Var:\ \ "\isasymlceil Var\ x\isasymrceil \ =\ SUCC\ (ORD\_OF\ (nat\_of\_name\ x))"\isanewline
\isacommand{lemma}\ quot\_Eats:\ "\isasymlceil Eats\ x\ y\isasymrceil \ =\ HPair\ (HTuple\ 1)\ (HPair\ \isasymlceil x\isasymrceil \ \isasymlceil y\isasymrceil )"\isanewline
\isacommand{lemma}\ quot\_Eq:\ \ \ "\isasymlceil x\ EQ\ y\isasymrceil \ =\ HPair\ (HTuple\ 2)\ (HPair\ (\isasymlceil x\isasymrceil )\ (\isasymlceil y\isasymrceil ))"\isanewline
\isacommand{lemma}\ quot\_Disj:\ "\isasymlceil A\ OR\ B\isasymrceil \ =\ HPair\ (HTuple\ 3)\ (HPair\ (\isasymlceil A\isasymrceil )\ (\isasymlceil B\isasymrceil ))"\isanewline
\isacommand{lemma}\ quot\_Ex:\ \ \ "\isasymlceil Ex\ i\ A\isasymrceil \ =\ HPair\ (HTuple\ 5)\ (quot\_dbfm\ (trans\_fm\ [i]\ A))"
\end{isabelle}
Note that \isa{HPair} constructs an HF term denoting a pair, while \isa{HTuple}~$n$ constructs 
an $(n+2)$-tuple of zeros.
Proofs often refer to the denotations of terms rather than to the terms themselves,
so the functions \isa{q\_Eats}, \isa{q\_Mem}, \isa{q\_Eq}, \isa{q\_Neg}, \isa{q\_Disj}, \isa{q\_Ex}
are defined to express these codes. Here are some examples:
\begin{isabelle}
\ \ \ "q\_Var\ i\ \isasymequiv \ succ\ (ord\_of\ (nat\_of\_name\ i))"\isanewline
\ \ \ "q\_Eats\ x\ y\ \isasymequiv \ \isasymlangle htuple\ 1,\ x,\ y\isasymrangle "\isanewline
\ \ \ "q\_Disj\ x\ y\ \isasymequiv \ \isasymlangle htuple\ 3,\ x,\ y\isasymrangle "\isanewline
\ \ \ "q\_Ex\ x\ \isasymequiv \ \isasymlangle htuple\ 5,\ x\isasymrangle "
\end{isabelle}
Note that \isa{\isasymlangle x,y\isasymrangle} denotes the pair of \isa{x} and \isa{y} as sets, in other words, of type \isa{hf}.

\subsection{Predicates for the Coding of Syntax}
\label{sec:Coding_Predicates}

The next and most arduous step is to define logical predicates corresponding to each of the syntactic concepts (terms, formulas, abstraction, substitution) mentioned above. Textbooks and articles describe each predicate at varying levels of detail. G\"odel \cite{goedel-formally} gives full definitions, as does {\'S}wierczkowski. Boolos \cite{boolos-provability} gives many details of how coding is set up, and gives the predicates for terms and formulas, but not for any operations upon them. Hodel \cite{hodel-intro}, like many textbook authors, relies heavily on ``algorithms'' written in English. The definitions indeed amount to pages of computer code. Authors typically conclude with a ``theorem'' asserting the correctness of this code. For example, {\'S}wierczkowski \cite[Sect.\ts3--4]{swierczkowski-finite} presents 34 highly technical definitions, justified by seven lines of proof.

Proving the correctness of this lengthy series of definitions requires a substantial effort,
and the proofs (being syntactically oriented) are tiresome.
It is helpful to introduce shadow versions of all predicates
in Isabelle/HOL's native logic, as well as in HF\@.
Having two versions of each predicate simplifies the task of relating the HF version of the predicate
to the syntactic concept that it is intended to represent;
the first step is to prove that the HF formula is equivalent to the syntactically similar definition written
in Isabelle's higher-order logic, which then is proved to satisfy deeper properties.
The shadow predicates also give an easy way to refer to the truth of the corresponding HF predicate;
because each is defined to be a $\Sigma$ formula, that gives a quick way
(using theorem \isa{Sigma\_fm\_imp\_thm} above) to show that some ground instance of the predicate can be proved formally in HF\@. Also, one way to arrive at the correct definition of an HF predicate is to define its shadow equivalent first, since proving that it implies the required properties is much easier in Isabelle/HOL's native logic than in HF\@.

{\'S}wierczkowski \cite{swierczkowski-finite}
defines a full set of syntactic predicates, leaving nothing as an exercise.
Unfortunately, the introduction of de Bruijn syntax necessitates rewriting many of these definitions. 
Some predicates (such as the variable occurrence test)
are replaced by others (abstraction over a variable occurrence).
The final list includes predicates to recognise the following items:
\begin{itemize}
\item the codes of well-formed terms (and constant terms, without variables)
\item correct instances of abstraction (of a term or formula) over a variable
\item correct instances of substitution (in a term or formula) for a variable
\item the codes of well-formed formulas
\end{itemize}
As explained below, abstraction over a formula needs to be defined before the notion of a formula itself.
We also need the property of variable non-occurrence, ``$x$ does not occur in~$A$'', which can be expressed directly as ``substituting 0 for $x$ in $A$ yields~$A$''. This little trick eliminates the need for a full definition.

Each operation is first defined in its sequence form (expressing that sequence $s$
is built up in an appropriate way and that $s_k$ is a specific value); existential quantification over $s$ and $k$ then yields the final predicate. Formalising the sequence of steps is a primitive way to express recursion. Moreover, it tends to yield $\Sigma$ formulas.

The simplest example is the predicate for constants. The shadow predicate can be defined with the help
of \isa{BuildSeq}, mentioned in Sect.\ts\ref{sec:functions} above. Note that shadow predicates
are written in ordinary higher-order logic, and refer to syntactic codes using set values.
We see below that in the sequence buildup, each element is either 0 (which is the code of the constant zero)
or else has the form \isa{q\_Eats\ v\ w}, which is the code for~$v\lhd w$.
\begin{isabelle}
\isacommand{definition}\ SeqConst\ ::\ "hf\ \isasymRightarrow \ hf\ \isasymRightarrow \ hf\ \isasymRightarrow \ bool"\isanewline
\ \ \isakeyword{where}\ "SeqConst\ s\ k\ t\ \isasymequiv \ BuildSeq\ (\isasymlambda u.\ u=0)\ (\isasymlambda u\ v\ w.\ u\ =\ q\_Eats\ v\ w)\ s\ k\ t"
\end{isabelle}
Thus a constant expression is built up from 0 using the $\lhd$ operator.  The idea is that every element of the sequence is either 0 or has the form $\quot{x\lhd y}$, where $x$ and $y$ occur earlier in the sequence. Most of the other syntactic predicates fit exactly the same pattern, but with different base cases and constructors.  A function must be coded as a relation, and a typical base case might be $\tuple{0,0}$, other sequence elements having the form $\tuple{\quot{x\lhd y}, \quot{x'\lhd y'}}$, where $\tuple{x,x'}$ and $\tuple{y,y'}$ occur earlier in the sequence. Substitution is codified in this manner. A function taking two arguments is coded as a sequence of triples, etc. 

The discussion above relates to shadow predicates, which define formulas of Isabelle/HOL relating HF sets. The real predicates, which denote formulas of the HF calculus, are based on exactly the same ideas except that the various set constructions must be expressed using the HF term language. Note that the real predicates typically have names ending with P\@.

The following formula again specifies the notion of a constant term. It is simply the result of expressing the definition of \isa{SeqConst} using HF syntax, expanding the definition of \isa{BuildSeq}. The syntactic sugar for a reference to a sequence element $s_m$ within some formula $\phi$ must now be expanded to its true form: $\phi(s_m)$ becomes $\ex{y}{\tuple{m,y}\in s\land \phi(y)}$.
\begin{isabelle}
\isacommand{nominal\_primrec}\ SeqConstP\ ::\ "tm\ \isasymRightarrow \ tm\ \isasymRightarrow \ tm\ \isasymRightarrow \ fm"\isanewline
\ \ \isakeyword{where}\ "\isasymlbrakk atom\ l\ \isasymsharp \ (s,k,sl,m,n,sm,sn);\ \ atom\ sl\ \isasymsharp \ (s,m,n,sm,sn);\isanewline
\ \ \ \ \ \ \ \ \ \ \ atom\ m\ \isasymsharp \ (s,n,sm,sn);\ \ atom\ n\ \isasymsharp \ (s,sm,sn);\isanewline
\ \ \ \ \ \ \ \ \ \ \ atom\ sm\ \isasymsharp \ (s,sn);\ \ atom\ sn\ \isasymsharp \ (s)\isasymrbrakk \ \isasymLongrightarrow \isanewline
\ \ \ \ SeqConstP\ s\ k\ t\ =\isanewline
\ \ \ \ \ \ LstSeqP\ s\ k\ t\ AND\isanewline
\ \ \ \ \ \ All2\ l\ (SUCC\ k)\ (Ex\ sl\ (HPair\ (Var\ l)\ (Var\ sl)\ IN\ s\ AND\ (Var\ sl\ EQ\ Zero\ OR\isanewline
\ \ \ \ \ \ \ \ \ \ \ \ \ Ex\ m\ (Ex\ n\ (Ex\ sm\ (Ex\ sn\ (Var\ m\ IN\ Var\ l\ AND\ Var\ n\ IN\ Var\ l\ AND\isanewline
\ \ \ \ \ \ \ \ \ \ \ \ \ \ \ \ \ HPair\ (Var\ m)\ (Var\ sm)\ IN\ s\ AND\ HPair\ (Var\ n)\ (Var\ sn)\ IN\ s\ AND\isanewline
\ \ \ \ \ \ \ \ \ \ \ \ \ \ \ \ \ Var\ sl\ EQ\ Q\_Eats\ (Var\ sm)\ (Var\ sn))))))))"
\end{isabelle}
We have five bound variables, namely \isa{l}, \isa{sl}, \isa{m}, \isa{sm}, \isa{n}, \isa{sn},
which must be constrained to be distinct from one another using the freshness conditions shown. This nominal boilerplate may seem excessive. However, to define this predicate without nominal syntax, bound variable names might have to be calculated, perhaps by taking the maximum of the bound variables in  \isa{s}, \isa{k} and \isa{t} and continuing from there. Nominal constrains the variables more abstractly and flexibly.

As mentioned above, sometimes we deal with sequences of pairs or triples, with correspondingly more complicated formulas. Where a predicate describes a function such as substitution, the sequence being built up consists of ordered pairs of arguments and results, and there are typically nine bound variables. To perform abstraction over a formula requires keeping track of the depth of quantifier nesting during recursion. This is a two-argument function, so the sequence being built up consists of ordered triples and there are 12 bound variables. Although the nominal system copes with these complicated expressions, the processing time can be measured in tens of seconds.

Now that we have defined the buildup of a sequence of constants, a constant itself is trivial. The existence of any properly formed sequence \isa{s} of length \isa{k} culminating with some term \isa{t} guarantees that \isa{t}  is a constant term.
Here are both the shadow and HF calculus versions of the predicate.
\begin{isabelle}
\isacommand{definition}\ Const\ ::\ "hf\ \isasymRightarrow \ bool"\isanewline
\ \ \isakeyword{where}\ "Const\ t\ \isasymequiv \ (\isasymexists s\ k.\ SeqConst\ s\ k\ t)"\vskip1ex
\isacommand{nominal\_primrec}\ ConstP\ ::\ "tm\ \isasymRightarrow \ fm"\isanewline
\ \ \isakeyword{where}\ "\isasymlbrakk atom\ k\ \isasymsharp \ (s,t);\ atom\ s\ \isasymsharp \ t\isasymrbrakk \ \isasymLongrightarrow \isanewline
\ \ \ \ ConstP\ t\ =\ Ex\ s\ (Ex\ k\ (SeqConstP\ (Var\ s)\ (Var\ k)\ t))"
\end{isabelle}

Why don't we define the HF predicate \isa{BuildSeqP}
analogously to \isa{BuildSeq}, which expresses the definition of \isa{SeqConst}
so succinctly? 
Then we might expect to avoid the mess above, defining a predicate such as \isa{SeqConstP} in a single line.
This was actually attempted, but the nominal system is not really suitable for formalising
higher-order definitions. Complicated auxiliary definitions and proofs are required.
It is easier simply to write out the definitions, especially as their very repetitiveness
allows proof development by cut and paste.

One tiny consolidation has been done. We need to define the predicates \isa{Term} and \isa{TermP}  analogously to \isa{Const} and \isa{ConstP} above but allowing variables.
The question of whether variables are allowed in a term or not can be governed by a Boolean.
The proof development therefore introduces the predicate \isa{SeqCTermP}, taking a Boolean argument,
from which \isa{SeqTermP} and \isa{SeqConstP} are trivially obtained.
\begin{isabelle}
\isacommand{abbreviation}\ SeqTermP\ ::\ "tm\ \isasymRightarrow \ tm\ \isasymRightarrow \ tm\ \isasymRightarrow \ fm"\isanewline
\ \ \isakeyword{where}\ "SeqTermP\ \isasymequiv \ SeqCTermP\ True"\isanewline
\isanewline
\isacommand{abbreviation}\ SeqConstP\ ::\ "tm\ \isasymRightarrow \ tm\ \isasymRightarrow \ tm\ \isasymRightarrow \ fm"\isanewline
\ \ \isakeyword{where}\ "SeqConstP\ \isasymequiv \ SeqCTermP\ False"
\end{isabelle}
In this way, we can define the very similar predicates \isa{TermP} and \isa{ConstP}
with a minimum of repeated material.

Many other predicates must be defined. 
Abstraction and substitution must be defined separately for terms, atomic formulas and formulas. 
Here are the shadow definitions of abstraction and substitution for terms. They are similar enough (both involve replacing a variable) that a single function, \isa{SeqStTerm}, can express both. \isa{BuildSeq2} is similar to \isa{BuildSeq} above, but constructs a sequence of pairs.
\begin{isabelle}
\isacommand{definition}\ SeqStTerm\ ::\ "hf\ \isasymRightarrow \ hf\ \isasymRightarrow \ hf\ \isasymRightarrow \ hf\ \isasymRightarrow \ hf\ \isasymRightarrow \ hf\ \isasymRightarrow \ bool"\isanewline
\isakeyword{where}\ "SeqStTerm\ v\ u\ x\ x'\ s\ k\ \isasymequiv \isanewline
\ \ \ is\_Var\ v\ \isasymand \ BuildSeq2\ (\isasymlambda y\ y'.\ (is\_Ind\ y\ \isasymor \ Ord\ y)\ \isasymand \ y'\ =\ (if\ y=v\ then\ u\ else\ y))\isanewline
\ \ \ \ \ \ \ \ \ \ \ \ \ \ \ \ (\isasymlambda u\ u'\ v\ v'\ w\ w'.\ u\ =\ q\_Eats\ v\ w\ \isasymand \ u'\ =\ q\_Eats\ v'\ w')\ s\ k\ x\ x'"\isanewline
\isanewline
\isacommand{definition}\ AbstTerm\ ::\ "hf\ \isasymRightarrow \ hf\ \isasymRightarrow \ hf\ \isasymRightarrow \ hf\ \isasymRightarrow \ bool"\isanewline
\ \ \isakeyword{where}\ "AbstTerm\ v\ i\ x\ x'\ \isasymequiv \ Ord\ i\ \isasymand \ (\isasymexists s\ k.\ SeqStTerm\ v\ (q\_Ind\ i)\ x\ x'\ s\ k)"\isanewline\isanewline
\isacommand{definition}\ SubstTerm\ ::\ "hf\ \isasymRightarrow \ hf\ \isasymRightarrow \ hf\ \isasymRightarrow \ hf\ \isasymRightarrow \ bool"\isanewline
\ \ \isakeyword{where}\ "SubstTerm\ v\ u\ x\ x'\ \isasymequiv \ Term\ u\ \isasymand \ (\isasymexists s\ k.\ SeqStTerm\ v\ u\ x\ x'\ s\ k)"
\end{isabelle}

Abstraction over formulas (\isa{AbstForm}/\isa{AbstFormP})
must be defined before formulas themselves, in order to 
formalise existential quantification. There is no circularity here: the abstraction operation
can be defined independently of the notion of a well-formed formula, and is not restricted to them. The definition involves sequences of triples and is too complicated to present here. 

With abstraction over formulas defined, we can finally define the concept of a formula itself. An \isa{Atomic} formula involves two terms,
combined using the relations \isa{EQ} or \isa{IN}\@.
Then \isa{MakeForm} combines one or two existing formulas to build more complex ones.
It constrains \isa{y} to be the code of a formula constructed from existing formulas \isa{u} and \isa{v}
by the disjunction $\isa{u} \lor \isa{v}$, the negation $\neg\isa{u}$ or the
existential formula $\exists(\isa{u'})$, where \isa{u'} has been obtained by abstracting \isa{u} over some variable, \isa{v} via the predicate \isa{AbstForm}.
\begin{isabelle}
\isacommand{definition}\ MakeForm\ ::\ "hf\ \isasymRightarrow \ hf\ \isasymRightarrow \ hf\ \isasymRightarrow \ bool"\isanewline
\ \ \isakeyword{where}\ "MakeForm\ y\ u\ w\ \isasymequiv \isanewline
\ \ \ \ \ \ \ \ \ \ \ \ y\ =\ q\_Disj\ u\ w\ \isasymor \ y\ =\ q\_Neg\ u\ \isasymor \isanewline
\ \ \ \ \ \ \ \ \ \ \ \ (\isasymexists v\ u'.\ AbstForm\ v\ 0\ u\ u'\ \isasymand \ y\ =\ q\_Ex\ u')"\isanewline
\isanewline
\isacommand{nominal\_primrec}\ MakeFormP\ ::\ "tm\ \isasymRightarrow \ tm\ \isasymRightarrow \ tm\ \isasymRightarrow \ fm"\isanewline
\ \ \isakeyword{where}\ "\isasymlbrakk atom\ v\ \isasymsharp \ (y,u,w,au);\ atom\ au\ \isasymsharp \ (y,u,w)\isasymrbrakk \ \isasymLongrightarrow \isanewline
\ \ \ \ MakeFormP\ y\ u\ w\ =\isanewline
\ \ \ \ \ \ y\ EQ\ Q\_Disj\ u\ w\ OR\ y\ EQ\ Q\_Neg\ u\ OR\isanewline
\ \ \ \ \ \ Ex\ v\ (Ex\ au\ (AbstFormP\ (Var\ v)\ Zero\ u\ (Var\ au)\ AND\ y\ EQ\ Q\_Ex\ (Var\ au)))"
\end{isabelle}
Now, the sequence buildup of a formula can be defined with \isa{Atomic} covering the base case and \isa{MakeForm} expressing one level of the construction. Using similar methods to those illustrated above for constant terms, we arrive at the shadow predicate \isa{Form} and the corresponding HF predicate \isa{FormP}.

\subsection{Verifying the Coding Predicates}

Most textbook presentations take the correctness of such definitions as obvious,
and indeed many properties are not difficult to prove.
To show that the predicate \isa{Term} accepts all coded terms, a necessary lemma
is to show the analogous property for well-formed de Bruijn terms:
\begin{isabelle}
\isacommand{lemma}\ wf\_Term\_quot\_dbtm:\isanewline
\ \ \isakeyword{assumes}\ "wf\_dbtm\ u"\ \isakeyword{shows}\ "Term\ \isasymlbrakk quot\_dbtm\ u\isasymrbrakk e"
\end{isabelle}
The proof is by induction on the construction of \isa{u} 
(in other words, on the inductive definition of \isa{wf\_dbtm\ u}),
and is routine by the definitions of the predicates \isa{Term} and \isa{SeqTerm}.
This implies the desired result for terms, by the (overloaded)
definition of \isa{\isasymlceil t\isasymrceil}.
\begin{isabelle}
\isacommand{corollary}\ Term\_quot\_tm:\ \isakeyword{fixes}\ t::tm\ \ \isakeyword{shows}\ "Term\ \isasymlbrakk\isasymlceil t\isasymrceil\isasymrbrakk e"
\end{isabelle}
Note that both results concern the shadow predicate \isa{Term}, not the HF predicate \isa{TermP}\@.
The argument of \isa{Term} is a set, denoted using the evaluation operator, \isa{\isasymlbrakk\ldots\isasymrbrakk e}. Direct proofs about HF predicates are long and tiresome. Fortunately, many such questions can be reduced to the corresponding questions involving shadow predicates, because codes are ground terms; then the theorem \isa{Sigma\_fm\_imp\_thm} guarantees the existence of a proof, sparing us the need to construct one.

The converse correctness property must also be proved, namely that everything accepted by \isa{Term}
actually is the code of some term. The proof requires a lemma
about the predicate \isa{SeqTerm}. The reasoning is simply that constants and variables are well-formed, and that combining two well-formed terms preserves this property. Such proofs are streamlined through the use of \isa{BuildSeq\_induct}, a derived rule for reasoning about  sequence construction by induction on the length of the sequence.
\begin{isabelle}
\isacommand{lemma}\ Term\_imp\_wf\_dbtm:\isanewline
\ \ \isakeyword{assumes}\ "Term\ x"\ \isakeyword{obtains}\ t::dbtm\ \isakeyword{where}\ "wf\_dbtm\ t"\ "x\ =\ \isasymlbrakk quot\_dbtm\ t\isasymrbrakk e"
\end{isabelle}
By the meaning of \isakeyword{obtains}, we see that if \isa{Term\ x}
then there exists some well-formed de Bruijn term~\isa{t} whose code evaluates to~\isa{x}.
Since for every well-formed de Bruijn term there exists an equivalent standard term of type~\isa{tm},
we can conclude that \isa{Term\ x} implies that \isa{x} is the code of some term.
\begin{isabelle}
\isacommand{corollary}\ Term\_imp\_is\_tm:\isanewline
\ \ \isakeyword{assumes}\ "Term\ x"\ \isakeyword{obtains}\ t::tm\ \isakeyword{where}\ "x\ =\ \isasymlbrakk \isasymlceil t\isasymrceil \isasymrbrakk \ e"
\end{isabelle}

Similar theorems---with similar proofs---are necessary for each of the syntactic predicates. 
For example, the following result expresses that \isa{SubstForm} correctly models the result \isa{A(i::=t)} of 
substituting \isa{t} for \isa{i} in the formula~\isa{A}.
\begin{isabelle}
\isacommand{lemma}\ SubstForm\_quot\_unique:\ \isanewline
\ \ "SubstForm\ (q\_Var\ i)\ \isasymlbrakk \isasymlceil t\isasymrceil \isasymrbrakk e\ \isasymlbrakk \isasymlceil A\isasymrceil \isasymrbrakk e\ x'\ \isasymlongleftrightarrow \ x'\ =\ \isasymlbrakk \isasymlceil A(i::=t)\isasymrceil \isasymrbrakk e"
\end{isabelle}

\subsection{Predicates for the Coding of Deduction}\label{sec:deduction}

On the whole, the formalisation of deduction is quite different from the formalisation of syntactic 
operations, which mostly involve simulated recursion over terms or formulas. A proof step in the HF calculus
is an axiom, an axiom scheme or an inference rule.
Axioms and propositional inference rules are straightforward to recognise using the
existing syntactic primitives. Simply \isa{x\ EQ\ \isasymlceil refl\_ax\isasymrceil} tests
whether \isa{x} denotes the reflexivity axiom. More complicated are inference rules involving
quantification, where several syntactic conditions including abstraction and substitution
need to be checked in sequence. For example, consider \emph{specialisation axioms} of the form 
$\phi(t/x)\to\exists x\,\phi$.
\begin{isabelle}
\isacommand{nominal\_primrec}\ Special\_axP\ ::\ "tm\ \isasymRightarrow \ fm"\ \isakeyword{where}\isanewline
\ \ "\isasymlbrakk atom\ v\ \isasymsharp \ (p,sx,y,ax,x);\ atom\ x\ \isasymsharp \ (p,sx,y,ax);\isanewline
\ \ \ \ atom\ ax\ \isasymsharp \ (p,sx,y);\ atom\ y\ \isasymsharp \ (p,sx);\ atom\ sx\ \isasymsharp \ p\isasymrbrakk \ \isasymLongrightarrow \isanewline
\ \ Special\_axP\ p\ =\ Ex\ v\ (Ex\ x\ (Ex\ ax\ (Ex\ y\ (Ex\ sx\isanewline
\ \ \ \ \ \ \ \ \ \ \ \ \ \ \ \ \ \ \ (FormP\ (Var\ x)\ AND\ VarP\ (Var\ v)\ AND\ TermP\ (Var\ y)\ AND\isanewline
\ \ \ \ \ \ \ \ \ \ \ \ \ \ \ \ \ \ \ \ AbstFormP\ (Var\ v)\ Zero\ (Var\ x)\ (Var\ ax)\ AND\isanewline
\ \ \ \ \ \ \ \ \ \ \ \ \ \ \ \ \ \ \ \ SubstFormP\ (Var\ v)\ (Var\ y)\ (Var\ x)\ (Var\ sx)\ AND\isanewline
\ \ \ \ \ \ \ \ \ \ \ \ \ \ \ \ \ \ \ \ p\ EQ\ Q\_Imp\ (Var\ sx)\ (Q\_Ex\ (Var\ ax)))))))"
\end{isabelle}
This definition states that a specialisation axiom is created from a formula \isa{x}, a variable \isa{v} and a term~\isa{y}, combined by appropriate abstraction and substitution operations. Correctness means proving that this predicate exactly characterises the elements of the set \isa{special\_axioms}, which was used to define the HF calculus. The most complicated such scheme is the induction axiom HF3 (defined in Sect.\ts\ref{sec:hf-set-theory}),
with its three quantifiers. The treatment of the induction axiom requires nearly 180 lines,
of which 120 are devoted to proving correctness with respect to the HF calculus.

A proof of a theorem \isa{y} is a sequence \isa{s} of axioms and inference rules, ending with~\isa{y}:
\begin{isabelle}
\isacommand{definition}\ Prf\ ::\ "hf\ \isasymRightarrow \ hf\ \isasymRightarrow \ hf\ \isasymRightarrow \ bool"\isanewline
\ \ \isakeyword{where}\ "Prf\ s\ k\ y\ \isasymequiv \ BuildSeq\ (\isasymlambda x.\ x\ \isasymin \ Axiom)\isanewline
\ \ \ \ \ \ \ \ \ \ \ \ \ \ \ \ \ \ \ \ \ \ \ \ \ \ \ \ (\isasymlambda u\ v\ w.\ ModPon\ v\ w\ u\ \isasymor \ Exists\ v\ u\ \isasymor \ Subst\ v\ u)\ s\ k\ 
\end{isabelle}

Finally, \isa{y} codes a theorem provided it has a proof:
\begin{isabelle}
\isacommand{definition}\ Pf\ ::\ "hf\ \isasymRightarrow \ bool"\isanewline
\ \ \isakeyword{where}\ "Pf\ y\ \isasymequiv \ (\isasymexists s\ k.\ Prf\ s\ k\ y)"
\end{isabelle}

Having proved the correctness of the predicates formalising the axioms and rules, 
the correctness of \isa{Pf} follows easily. (\'Swierczkowski's seven lines of proof start here.)
One direction is proved by induction on the proof of \isa{\isacharbraceleft \isacharbraceright \ \isasymturnstile \ \isasymalpha}.
\begin{isabelle}
\isacommand{lemma}\ proved\_imp\_Pf:\ \isakeyword{assumes}\ "H\ \isasymturnstile \ \isasymalpha "\ "H=\isacharbraceleft \isacharbraceright "\ \isakeyword{shows}\ "Pf\ \isasymlbrakk \isasymlceil \isasymalpha \isasymrceil \isasymrbrakk e"
\end{isabelle}
Here, we use the shadow predicates and work directly in Isabelle/HOL\@. The corresponding HF predicate, \isa{PfP}, is (crucially) a $\Sigma$ formula. Moreover, codes are ground terms. Therefore \isa{PfP\ \isasymlceil \isasymalpha \isasymrceil} is a $\Sigma$ sentence and is formally provable. This is the main use of the theorem \isa{Sigma\_fm\_imp\_thm}.
\begin{isabelle}
\isacommand{corollary}\ proved\_imp\_proved\_PfP:\ "\isacharbraceleft \isacharbraceright \ \isasymturnstile \ \isasymalpha \ \isasymLongrightarrow \ \isacharbraceleft \isacharbraceright \ \isasymturnstile \ PfP\ \isasymlceil \isasymalpha \isasymrceil "
\end{isabelle}
The reverse implication, despite its usefulness, is not always proved. Again using the rule \isa{BuildSeq\_induct}, it holds by induction on the length of the coded proof of \isa{\isasymlceil \isasymalpha \isasymrceil}. The groundwork for this result involves proving, for each coded axiom and inference rule, that there exists a corresponding proof step in the HF calculus. 
We continue to work at the level of shadow predicates. 
\begin{isabelle}
\isacommand{lemma}\ Prf\_imp\_proved:\ \isakeyword{assumes}\ "Prf\ s\ k\ x"\ \isakeyword{shows}\ "\isasymexists A.\ x\ =\ \isasymlbrakk \isasymlceil A\isasymrceil \isasymrbrakk e\ \isasymand \ \isacharbraceleft \isacharbraceright \ \isasymturnstile \ A"
\end{isabelle}
The corresponding result for \isa{Pf} is immediate:
\begin{isabelle}
\isacommand{corollary}\ Pf\_quot\_imp\_is\_proved:\ "Pf\ \isasymlbrakk \isasymlceil \isasymalpha \isasymrceil \isasymrbrakk e\ \isasymLongrightarrow \ \isacharbraceleft \isacharbraceright \ \isasymturnstile \ \isasymalpha "
\end{isabelle}
Now \isa{\isacharbraceleft \isacharbraceright \ \isasymturnstile \ PfP\ \isasymlceil \isasymalpha \isasymrceil} implies 
\isa{Pf\ \isasymlbrakk \isasymlceil \isasymalpha \isasymrceil \isasymrbrakk e}
simply by the soundness of the calculus. It is now easy to show that the predicate \isa{PfP} corresponds exactly to deduction in the HF calculus.
\'Swierczkowski calls this result the \emph{proof formalisation condition}.
\begin{isabelle}
\isacommand{theorem}\ proved\_iff\_proved\_PfP:\ "\isacharbraceleft \isacharbraceright \ \isasymturnstile \ \isasymalpha \ \isasymlongleftrightarrow \ \isacharbraceleft \isacharbraceright \ \isasymturnstile \ PfP\ \isasymlceil \isasymalpha \isasymrceil "
\end{isabelle}

\emph{Remark}: \isa{PfP} includes an additional primitive inference, substitution:
\[ \frac{H\vdash \alpha}{H\vdash \alpha(t/x)} \]
This inference is derivable in the HF calculus, but the 
second incompleteness theorem requires performing substitution inferences,
and reconstructing the derivation of substitution within \isa{PfP} would be infeasible.
Including substitution in the definition of \isa{PfP} makes such steps immediate
without complicating other proofs.
\'Swierczkowski avoids this complication: his HF calculus \cite{swierczkowski-finite} includes
a substitution rule alongside a simpler specialisation axiom.

\subsection{Pseudo-Functions}

The HF calculus contains no function symbols other than~$\lhd$. All other ``functions'' must be declared as predicates, which are mere abbreviations of formulas. This abuse of notation is understood in a standard way. The formula $\phi(f(x))$ can be taken as an abbreviation for $\ex{y}{F(x,y)\land \phi(y)}$
where $F$ is the relation representing the function $f$ and provided that $F$ can be proved to be single valued: ${F(x,y), F(x,y')}\vdash y'=y$. Then $f$ is called a \textit{pseudo-function} and something like $f(x)$ is called a \textit{pseudo-term}. Pseudo-terms do not actually exist, which will cause problems later.

G\"odel formalised all syntactic operations
as primitive recursive functions, while Boolos \cite{boolos-provability} used $\Delta$ formulas.
With both approaches, much effort is necessary to admit a function definition in the first place.
But then, it is known to be a function.
Here we see a drawback of \'Swierczkowski's decision to base the formalisation on the notion of $\Sigma$ formulas: they do not cover the property of being single valued. A predicate that corresponds to a function must be proved to be single valued within the HF calculus itself. G\"odel's proof uses  substitution as a function. The proof that substitution (on terms and formulas) is single valued requires nearly 500 lines
in Isabelle/HOL, not counting considerable preparatory material (such as the partial ordering properties of $<$)
mentioned in Sect.\ts\ref{sec:functions} above.

Fortunately, these proofs are conceptually simple and highly repetitious, and again much of the proof development can be done
by cut and paste. The first step is to prove an \emph{unfolding lemma} about the sequence buildup:
if the predicate holds, then either the base case holds, or else 
there exist values earlier in the sequence for which one of the recursive cases can be applied.
The single valued theorem is proved by complete induction on the length of the sequence,
with a fully quantified induction formula (analogous to $\all{x y y'}{F(x,y) \to F(x,y') \to y'=y}$)
so that the induction hypothesis says that all shorter sequences
are single valued for all possible arguments. All that is left is to
apply the unfolding lemma to both instances of the relation~$F$,
and then to consider all combinations of cases. Most will be trivially contradictory,
and in those few cases where the result has the same outer form, an appeal to the induction hypothesis
for the operands will complete the proof.

Overall, the G\"odel development proves single valued theorems for 12 predicates.
Five of the theorems are proved by induction as sketched above. Here is an example:
\begin{isabelle}
\isacommand{lemma}\ SeqSubstFormP\_unique:\isanewline
\ \ "\isacharbraceleft SeqSubstFormP\ v\ a\ x\ y\ s\ u,\ SeqSubstFormP\ v\ a\ x\ y'\ s'\ u'\isacharbraceright \ \isasymturnstile \ y'\ EQ\ y"
\end{isabelle}
The remaining results are straightforward corollaries of those inductions:
\begin{isabelle}
\isacommand{theorem}\ SubstFormP\_unique:\isanewline
\ \ "\isacharbraceleft SubstFormP\ v\ tm\ x\ y,\ SubstFormP\ v\ tm\ x\ y'\isacharbraceright \ \isasymturnstile \ y'\ EQ\ y"
\end{isabelle}
It is worth repeating that these proofs are formally conducted within the HF calculus, essentially by single-step inferences. Meta-theory is no help here.

\section{G\"odel's First Incompleteness Theorem} \label{sec:G-I}

Discussions involving encodings frequently need a way to refer to syntactic objects.
We often see the convention where if $x$ is a natural number, then a boldface $\boldsymbol{x}$ 
stands for the corresponding numeral. Then in expressions like 
$x=y\to\Pf{\quot{\boldsymbol{x}=\boldsymbol{y}}}$, 
we see that the boldface convention actually abbreviates the function $x\mapsto\boldsymbol{x}$, 
which needs to be
formalisable in the HF calculus.

Thus, we need to define a function $Q$ such that $Q(x)=\quot{x}$, in other words,
$Q(x)$ yields some term~$t$ whose denotation is~$x$. This is trivial
if $x$ ranges over the natural numbers, by primitive recursion. 
It is problematical when $x$ ranges over sets, because it requires a canonical ordering over
the universe of sets. We don't need to solve this problem just yet:
the \emph{first} incompleteness theorem needs
only a function $H$ such that $H(\quot{A})=\quot{\quot{A}}$ for all $A$.
Possible arguments of $H$ are not arbitrary sets, but only nested tuples of ordinals;
these have a canonical form, so a functional relationship is easy
to define \cite{swierczkowski-finite}. A certain amount of effort establishes the required property:%
\footnote{Here \isa{\isakeyword{fixes}\ A::fm} declares \isa{A} 
to be a formula in the overloaded notation \isa{\isasymlceil A\isasymrceil}.
\'Swierczkowski uses $\alpha$, $\beta$, \ldots{} to denote formulas, but I've frequently used the traditional $A$, $B$, \ldots{}.}
\begin{isabelle}
\isacommand{lemma}\ prove\_HRP:\ \isakeyword{fixes}\ A::fm\ \isakeyword{shows}\ "\isacharbraceleft \isacharbraceright \ \isasymturnstile \ HRP\ \isasymlceil A\isasymrceil \ \isasymlceil \isasymlceil A\isasymrceil \isasymrceil "
\end{isabelle}
Note that the function $H$ has been formalised as the relation \isa{HRP};
it is defined using sequence operators, \isa{LstSeqP}, etc., as we have seen already.

In order to prove G\"odel's diagonal lemma, we need a function $K_i$ to substitute the code of a formula into itself, replacing the variable $x_i$. This function, again, is realised as a relation,
combining \isa{HRP} with \isa{SubstFormP}.
\begin{isabelle}
\isacommand{nominal\_primrec}\ KRP\ ::\ "tm\ \isasymRightarrow \ tm\ \isasymRightarrow \ tm\ \isasymRightarrow \ fm"\isanewline
\ \ \isakeyword{where}\ "atom\ y\ \isasymsharp \ (v,x,x')\ \isasymLongrightarrow \isanewline
\ \ \ \ \ \ \ \ \ KRP\ v\ x\ x'\ =\ Ex\ y\ (HRP\ x\ (Var\ y)\ AND\ SubstFormP\ v\ (Var\ y)\ x\ x')"
\end{isabelle}
We easily obtain a key result: $K_i \,(\quot{\alpha}) =\quot{\alpha(i:=\quot{\alpha})}$.
\begin{isabelle}
\isacommand{lemma}\ prove\_KRP:\ "\isacharbraceleft \isacharbraceright \ \isasymturnstile \ KRP\ \isasymlceil Var\ i\isasymrceil \ \isasymlceil \isasymalpha\isasymrceil \ \isasymlceil \isasymalpha(i::=\isasymlceil \isasymalpha\isasymrceil )\isasymrceil"
\end{isabelle}
However, it is essential to prove that \isa{KRP} behaves like a function.
The predicates \isa{KRP} and \isa{HRP} can be proved
to be single valued using the techniques discussed in the previous section. 
Then an appeal to \isa{prove\_KRP} uniquely characterises $K_i$ as a function:
\begin{isabelle}
\isacommand{lemma}\ KRP\_subst\_fm:\ "\isacharbraceleft KRP\ \isasymlceil Var\ i\isasymrceil \ \isasymlceil \isasymalpha \isasymrceil \ (Var\ j)\isacharbraceright \ \isasymturnstile \ Var\ j\ EQ\ \isasymlceil \isasymalpha (i::=\isasymlceil \isasymalpha \isasymrceil )\isasymrceil "
\end{isabelle}

Twenty five lines of tricky reasoning are needed to reach the next milestone: the \emph{diagonal lemma}.
\'Swierczkowski writes
\begin{quote}
We replace the variable $x_i$ in $\alpha$ by the [pseudo-term $K_i(x_i)$], and we denote by $\beta$ the resulting formula.
\cite[p.\ts22]{swierczkowski-finite}
\end{quote}
The elimination of the pseudo-function $K_i$ in favour of an existential quantifier involving \isa{KRP}
yields the following not-entirely-obvious Isabelle definition:
\begin{isabelle}
\ \ \isacommand{def}\ \isasymbeta \ \isasymequiv \ "Ex\ j\ (KRP\ \isasymlceil Var\ i\isasymrceil \ (Var\ i)\ (Var\ j)\ AND\ \isasymalpha (i\ ::=\ Var\ j))"
\end{isabelle}
Note that one occurrence of \isa{Var i} is quoted and the other is not. The development is full of pitfalls such as these.

The statement of the diagonal lemma is as follows. 
The second assertion states that the free variables of \isa{\isasymdelta}, the diagonal formula,
are those of \isa{\isasymalpha}, the original formula, minus~\isa{i}.
\begin{isabelle}
\isacommand{lemma}\ diagonal:\ \isanewline
\ \ \isakeyword{obtains}\ \isasymdelta \ \isakeyword{where}\ "\isacharbraceleft \isacharbraceright \ \isasymturnstile \ \isasymdelta \ IFF\ \isasymalpha (i::=\isasymlceil \isasymdelta \isasymrceil )"\ \ "supp\ \isasymdelta \ =\ supp\ \isasymalpha \ -\ \isacharbraceleft atom\ i\isacharbraceright "
\end{isabelle}

Figure~\ref{fig:G-I} presents the proof of the first incompleteness theorem itself.
The top level argument is quite simple, given the diagonal lemma. The key steps of the proof
should be visible even to somebody who is not an Isabelle expert, thanks to the structured Isar language.
Note that if \isa{\isacharbraceleft \isacharbraceright \ \isasymturnstile \ Neg\ \isasymdelta},
then \isa{\isacharbraceleft \isacharbraceright \ \isasymturnstile \ PfP \isasymlceil \isasymdelta \isasymrceil} and therefore
\isa{\isacharbraceleft \isacharbraceright \ \isasymturnstile\ \isasymdelta}
by the proof formalisation condition, violating the assumption of consistency.

\begin{figure}
\begin{isabelle}
\isacommand{theorem}\ Goedel\_I:\isanewline
\ \ \isakeyword{assumes}\ "\isasymnot \ \isacharbraceleft \isacharbraceright \ \isasymturnstile \ Fls"\isanewline
\ \ \isakeyword{obtains}\ \isasymdelta \ \isakeyword{where}\ "\isacharbraceleft \isacharbraceright \ \isasymturnstile \ \isasymdelta \ IFF\ Neg\ (PfP\ \isasymlceil \isasymdelta \isasymrceil )"\ \ "\isasymnot \ \isacharbraceleft \isacharbraceright \ \isasymturnstile \ \isasymdelta "\ \ "\isasymnot \ \isacharbraceleft \isacharbraceright \ \isasymturnstile \ Neg\ \isasymdelta "\ \ \isanewline
\ \ \ \ \ \ \ \ \ \ \ \ \ \ \ \ \ \ \ "eval\_fm\ e\ \isasymdelta "\ \ "ground\_fm\ \isasymdelta "\isanewline
\isacommand{proof}\ -\isanewline
\ \ \isacommand{fix}\ i::name\isanewline
\ \ \isacommand{obtain}\ \isasymdelta \ \isakeyword{where}\ \ \ \ \ \ \ \ "\isacharbraceleft \isacharbraceright \ \isasymturnstile \ \isasymdelta \ IFF\ Neg\ ((PfP\ (Var\ i))(i::=\isasymlceil \isasymdelta \isasymrceil ))"\isanewline
\ \ \ \ \ \ \ \ \ \ \ \ \ \isakeyword{and}\ suppd:\ \ "supp\ \isasymdelta \ =\ supp\ (Neg\ (PfP\ (Var\ i)))\ -\ \isacharbraceleft atom\ i\isacharbraceright "\isanewline
\ \ \ \ \isacommand{by}\ (metis\ SyntaxN.Neg\ diagonal)\isanewline
\ \ \isacommand{then}\ \isacommand{have}\ diag:\ "\isacharbraceleft \isacharbraceright \ \isasymturnstile \ \isasymdelta \ IFF\ Neg\ (PfP\ \isasymlceil \isasymdelta \isasymrceil )"\isanewline
\ \ \ \ \isacommand{by}\ simp\isanewline
\ \ \isacommand{then}\ \isacommand{have}\ np:\ "\isasymnot \ \isacharbraceleft \isacharbraceright \ \isasymturnstile \ \isasymdelta \ \isasymand \ \isasymnot \ \isacharbraceleft \isacharbraceright \ \isasymturnstile \ Neg\ \isasymdelta "\isanewline
\ \ \ \ \isacommand{by}\ (metis\ Iff\_MP\_same\ NegNeg\_D\ Neg\_D\ Neg\_cong\ assms\ proved\_iff\_proved\_PfP)\isanewline
\ \ \isacommand{then}\ \isacommand{have}\ "eval\_fm\ e\ \isasymdelta "\ \isacommand{using}\ hfthm\_sound\ [\isakeyword{where}\ e=e,\ OF\ diag]\isanewline
\ \ \ \ \isacommand{by}\ simp\ (metis\ Pf\_quot\_imp\_is\_proved)\isanewline
\ \ \isacommand{moreover}\ \isacommand{have}\ "ground\_fm\ \isasymdelta "\ \isacommand{using}\ suppd\ \ \isanewline
\ \ \ \ \isacommand{by}\ (simp\ add:\ supp\_conv\_fresh\ ground\_fm\_aux\_def\ subset\_eq)\isanewline
\ \ \ \ \ \ \ (metis\ fresh\_ineq\_at\_base)\isanewline
\ \ \isacommand{ultimately}\ \isacommand{show}\ ?thesis\isanewline
\ \ \ \ \isacommand{by}\ (metis\ diag\ np\ that)\isanewline
\isacommand{qed}%
\end{isabelle}
\caption{G\"odel's First Incompleteness Theorem}
\label{fig:G-I}
\end{figure}

\section{Towards the Second Theorem: Pseudo-Coding and Quotations} \label{sec:pseudo}

The second incompleteness theorem \cite{bagaria-second} has always been more mysterious than the first.
G\"odel's original paper \cite{goedel-formally} was designated ``Part I'' in anticipation of
a subsequent ``Part II'' proving the second theorem, but no second paper appeared.
Logicians recognised that the second theorem followed from the first, assuming that the first could itself be formalised in the internal calculus. While this assumption seems to be widely accepted, conducting such a formalisation explicitly remains difficult, even given today's theorem-proving technology.

A simpler route to the theorem involves the \emph{Hilbert-Bernays
derivability conditions} \cite[p.\ts15]{boolos-provability}\cite[p.\ts73]{grandy-advanced}.
\begin{align}
\text{If } \vdash \alpha\text{, then } &\vdash \Pf(\quot{\alpha})
  \tag{D1} \\
\text{If } \vdash \Pf(\quot{\alpha\to\beta})\text{ and } \vdash\Pf(\quot{\alpha})
\text{, then } &\vdash \Pf(\quot{\beta})  \tag{D2} \\
\text{If $\alpha$ is a strict $\Sigma$ sentence, then } &\vdash \alpha \to \Pf(\quot{\alpha})  \tag{D3} 
\end{align}
(Where there is no ambiguity, we identify Pf with the formalised predicate \isa{PfP};
the latter is the actual HF predicate, but the notation Pf is widely used in the literature,
along with G\"odel's original Bew.)

Condition (D1) is none other than the theorem \isa{proved\_iff\_proved\_PfP} mentioned 
in Sect.\ts\ref{sec:deduction} above. Condition (D2) seems clear by the definition of the predicate Pf, although all details of the workings of this predicate must be proved using low-level inferences in the HF calculus. Condition (D3) can be regarded as a version of the theorem \isa{Sigma\_fm\_imp\_thm} (``true $\Sigma$ sentences are theorems'') internalised as a theorem of the internal calculus. So while we avoid having to formalise the whole of G\"odel's theorem within the calculus, we end up formalising a key part of it.

Condition (D3) is stated in a general form, but we only need one specific instance:
\[ \vdash \Pf(\quot{\alpha}) \to \Pf(\quot{\Pf(\quot{\alpha})}). \]
Despite a superficial resemblance, (D3) does not follow from (D1), which holds by induction on the proof of $\vdash \alpha$.
As \'Swierczkowski explains \cite[p.\ts23]{swierczkowski-finite}, 
condition (D3) is not general enough to prove by induction.
In the sequel, we generalise and prove it.

\subsection{Pseudo-Coding}

Condition (D3) can be proved by induction on $\alpha$ if the assertion is generalised
so that $\alpha$ can have free variables, say $x_1$, \ldots,~$x_n$:
\[ \vdash \alpha(x_1,\ldots,x_n) \to \Pf(\quot{\alpha(\boldsymbol{x_1},\ldots,\boldsymbol{x_n})}) \]
The syntactic constructions used in this formula have to be formalised,
and the necessary transformations have to be justified within the HF calculus.
As mentioned above (Sect.\ts\ref{sec:G-I}), the boldface convention needs to be made rigorous.
In particular, codings are always ground HF terms, and yet $\quot{\alpha(\boldsymbol{x_1},\ldots,\boldsymbol{x_n})}$ has a functional dependence (as an HF term) on $x_1$, \ldots,~$x_n$.

The first step in this process is to generalise coding to allow certain variables to be preserved as variables in the coded term. Recall that with normal quotations, every occurrence of a variable is replaced by the code of the variable name, ultimately a positive integer:%
\footnote{\isa{ORD\_OF~(Suc~n)} denotes an HF term that denotes a positive integer, even if \isa{n} is a variable.}
\begin{isabelle}
\isacommand{function}\ quot\_dbtm\ ::\ "dbtm\ \isasymRightarrow \ tm"\isanewline
\ \ \isakeyword{where}\isanewline
\ \ \ "quot\_dbtm\ DBZero\ =\ Zero"\isanewline
\ |\ "quot\_dbtm\ (DBVar\ name)\ =\ ORD\_OF\ (Suc\ (nat\_of\_name\ name))"\isanewline
\ |\ \ldots
\end{isabelle}
Now let us define the \textit{$V$-code} of a term or formula, where $V$ is a set of variables
to be preserved in the code:
\begin{isabelle}
\isacommand{function}\ vquot\_dbtm\ ::\ "name\ set\ \isasymRightarrow \ dbtm\ \isasymRightarrow \ tm"\isanewline
\ \ \isakeyword{where}\isanewline
\ \ \ "vquot\_dbtm\ V\ DBZero\ =\ Zero"\isanewline
\ |\ "vquot\_dbtm\ V\ (DBVar\ name)\ =\ (if\ name\ \isasymin \ V\ then\ Var\ name\isanewline
\ \ \ \ \ \ \ \ \ \ \ \ \ \ \ \ \ \ \ \ \ \ \ \ \ \ \ \ \ \ \ \ \ else\ ORD\_OF\ (Suc\ (nat\_of\_name\ name)))"\isanewline
\ |\ \ldots
\end{isabelle}
$V$-coding is otherwise identical to standard coding, with the overloaded syntax
\isa{\isasymlfloor A\isasymrfloor V}.
The parameter $V$ is necessary because not all variables should be preserved;
it will be necessary to consider a series of $V$-codes for $V=\emptyset$, $\{x_1\}$ \ldots,
$\{x_1,\ldots,x_n\}$.

\subsection{Simultaneous Substitution}

In order to formalise the notation $\quot{\alpha(\boldsymbol{x_1},\ldots,\boldsymbol{x_n})}$,
it is convenient to introduce a function for simultaneous substitution.
Here \'Swierczkowski's presentation is a little hard to follow:
\begin{quote}
Suppose $\beta$ is a theorem, i.e., $\vdash\beta$. 
If we replace each of the variables at each of its free occurrences in~$\beta$ by some constant term
then the formula so obtained is also a theorem (by the Substitution Rule\ldots{}).
This just described situation in the meta-theory admits description in HF\@.
\cite[p.\ts24]{swierczkowski-finite}
\end{quote}
It took me weeks of failed attempts to grasp the meaning of the phrase ``constant term''. It does not mean a term containing no variables, but a term satisfying the predicate \isa{ConstP} and thus denoting the \emph{code} of a constant. Formalising this process seems to require replacing each variable $x_i$ by a new variable, $x'_i$,
intended to denote~$\boldsymbol{x_i}$. Later, it will be constrained to do so by a suitable HF predicate.
And so we need a function to perform simultaneous substitutions in a term
for all variables in a set~$V$. Using a ``fold'' operator over finite sets \cite{nipkow-defining} 
eliminates the need to consider the variables in any particular order.
\begin{isabelle}
\isacommand{definition}\ ssubst\ ::\ "tm\ \isasymRightarrow \ name\ set\ \isasymRightarrow \ (name\ \isasymRightarrow \ tm)\ \isasymRightarrow \ tm"\isanewline
\ \ \isakeyword{where}\ "ssubst\ t\ V\ F\ =\ Finite\_Set.fold\ (\isasymlambda i.\ subst\ i\ (F\ i))\ t\ V"
\end{isabelle}
The renaming of $x_i$ to~$x'_i$ could be done using arithmetic on variable subscripts,
but the formalisation instead follows an abstract approach, using nominal primitives.
An Isabelle locale defines a proof context containing a permutation~\isa{p} 
(mapping all variable names to new ones),
a finite set~\isa{Vs} of variable names and finally the actual renaming function~\isa{F},
which simply applies the permutation to any variable in~\isa{Vs}.%
\footnote{\isa{make\_F\ Vs\ p\ i\ =\ Var\ (p\ \isasymbullet \ i)} provided  \isa{i\ \isasymin \ Vs}.}
\begin{isabelle}
\isacommand{locale}\ quote\_perm\ =\isanewline
\ \ \isakeyword{fixes}\ p\ ::\ perm\ \isakeyword{and}\ Vs\ ::\ "name\ set"\ \isakeyword{and}\ F\ ::\ "name\ \isasymRightarrow \ tm"\isanewline
\ \ \isakeyword{assumes}\ p:\ "atom\ `\ (p\ \isasymbullet \ Vs)\ \isasymsharp *\ Vs"\ \isanewline
\ \ \ \ \ \ \isakeyword{and}\ pinv:\ "-p\ =\ p"\isanewline
\ \ \ \ \ \ \isakeyword{and}\ Vs:\ \ "finite\ Vs"\ \isanewline
\ \ \isakeyword{defines}\ "F\ \isasymequiv \ make\_F\ Vs\ p"
\end{isabelle}
Most proofs about \isa{ssubst} are done within the context of this locale, 
because it provides sufficient conditions
for the simultaneous substitution to be meaningful. The first condition states that \isa{p}
maps all the variables in \isa{Vs} to variables outside of that set,
while second condition states that \isa{p} is its own inverse.

This abstract approach is a little unwieldy at times, but its benefits can be seen
in the simple fact below, which states the effect of the simultaneous substitution on a single variable.
\begin{isabelle}
\isacommand{lemma}\ ssubst\_Var\_if:\isanewline
\ \ \isakeyword{assumes}\ "finite\ V"\ \ \isanewline
\ \ \ \ \isakeyword{shows}\ "ssubst\ (Var\ i)\ V\ F\ =\ (if\ i\ \isasymin \ V\ then\ F\ i\ else\ Var\ i)"
\end{isabelle}

We need to show that the variables in the set~\isa{Vs} can be renamed, one at a time, in a pseudo-coded de Bruijn term.
Let \isa{V\ \isasymsubseteq \ Vs} and
suppose that the variables in~\isa{V} have already been renamed, and choose one of the remaining
variables,~\isa{w}. It will be replaced by a new variable, computed by \isa{F~w}.
And something very subtle is happening: the variable \isa{w} is represented in the term
by its code, \isa{\isasymlceil Var\ w\isasymrceil}. 
Its replacement, \isa{F~w}, is \isa{Var\ (p\ \isasymbullet \ w)} and a \emph{variable}.
\begin{isabelle}
\isacommand{lemma}\ SubstTermP\_vquot\_dbtm:\isanewline
\ \ \isakeyword{assumes}\ w:\ "w\ \isasymin \ Vs - V"\ \isakeyword{and}\ V:\ "V\ \isasymsubseteq \ Vs"\ "V'\ =\ p\ \isasymbullet \ V"\isanewline
\ \ \ \ \ \ \isakeyword{and}\ s:\ "supp\ dbtm\ \isasymsubseteq \ atom\ `\ Vs"\isanewline
\ \ \isakeyword{shows}\isanewline
\ \ "insert\ (ConstP\ (F\ w))\ \isacharbraceleft ConstP\ (F\ i)\ |\ i.\ i\ \isasymin \ V\isacharbraceright \isanewline
\ \ \ \isasymturnstile \ SubstTermP\ \isasymlceil Var\ w\isasymrceil \ (F\ w)\isanewline
\ \ \ \ \ \ \ \ \ \ \ \ \ \ \ \ (ssubst\ (vquot\_dbtm\ V\ dbtm)\ V\ F)\isanewline
\ \ \ \ \ \ \ \ \ \ \ \ \ \ \ \ (subst\ w\ (F\ w)\ (ssubst\ (vquot\_dbtm\ (insert\ w\ V)\ dbtm)\ V\ F))"
\end{isabelle}
This theorem is proved by structural induction on \isa{dbtm}, the de Bruijn term.
The condition \isa{supp\ dbtm\ \isasymsubseteq \ atom\ `\ Vs} states that
the free variables of \isa{dbtm} all belong to~\isa{Vs}.
The variable case of the induction is tricky (and is the crux of the entire proof). We are working with a coded term that contains both coded variables and real ones (of the form \isa{F~i});
it is necessary to show that the real variables are preserved by the substitution,
because they are the $\boldsymbol{x_i}$ that we are trying to introduce.
The \isa{F~i} are preserved under the assumption that they denote constants,
which is the point of including the formulas 
\isa{ConstP\ (F\ i)} for all \isa{i\ \isasymin \ V}
on the left side of the turnstile. These assumptions will have to be justified later.

Under virtually the same assumptions (omitted), the analogous result holds for pseudo-coded de Bruijn formulas.
\begin{isabelle}
\isacommand{lemma}\ SubstFormP\_vquot\_dbfm:\isanewline
\ \ "insert\ (ConstP\ (F\ w))\ \isacharbraceleft ConstP\ (F\ i)\ |\ i.\ i\ \isasymin \ V\isacharbraceright \isanewline
\ \ \ \isasymturnstile \ SubstFormP\ \isasymlceil Var\ w\isasymrceil \ (F\ w)\isanewline
\ \ \ \ \ \ \ \ \ \ \ \ \ \ \ \ (ssubst\ (vquot\_dbfm\ V\ dbfm)\ V\ F)\ \isanewline
\ \ \ \ \ \ \ \ \ \ \ \ \ \ \ \ (subst\ w\ (F\ w)\ (ssubst\ (vquot\_dbfm\ (insert\ w\ V)\ dbfm)\ V\ F))"
\end{isabelle}
The proof is an easy structural induction on \isa{dbfm}. Every case holds immediately by properties of substitution and the induction hypotheses or by the previous theorem, for terms. The only difficult case in these two proofs is the variable case discussed above. Using the notation for $V$-coding, we can see that the substitution predicate \isa{SubstFormP}
can transform the term \isa{ssubst\ \isasymlfloor A\isasymrfloor V\ V\ F} into
\begin{quote}
\isa{ssubst\ \isasymlfloor A\isasymrfloor (insert\ w\ V)\ (insert\ w\ V)\ F}.
\end{quote}
Repeating this step, we can replace any finite set of variables in a coded formula by real ones, realising \'Swierczkowski's remark quoted at the top of this section, and in particular his last sentence. That is, if $\beta$ is a theorem in HF (if ${}\vdash\Pf\beta$ holds) then the result of substituting constants for its variables is also an HF theorem. More precisely still, we are replacing some subset \isa{V} of the variables by fresh variables (the \isa{F~i}), constrained by the predicate \isa{ConstP}.
\begin{isabelle}
\isacommand{theorem}\ PfP\_implies\_PfP\_ssubst:\isanewline
\ \ \isakeyword{fixes}\ \isasymbeta ::fm\isanewline
\ \ \isakeyword{assumes}\ \isasymbeta :\ "\isacharbraceleft \isacharbraceright \ \isasymturnstile \ PfP\ \isasymlceil \isasymbeta \isasymrceil "\isanewline
\ \ \ \ \ \ \isakeyword{and}\ V:\ "V\ \isasymsubseteq \ Vs"\isanewline
\ \ \ \ \ \ \isakeyword{and}\ s:\ "supp\ \isasymbeta \ \isasymsubseteq \ atom\ `\ Vs"\isanewline
\ \ \ \ \isakeyword{shows}\ \ \ \ "\isacharbraceleft ConstP\ (F\ i)\ |\ i.\ i\ \isasymin \ V\isacharbraceright \ \isasymturnstile \ PfP\ (ssubst\ \isasymlfloor \isasymbeta \isasymrfloor V\ V\ F)"
\end{isabelle}

The effort needed to formalise the results outlined above is relatively modest,
at 330 lines of Isabelle/HOL, but this excludes the effort needed to prove some essential lemmas,
which state that the various syntactic predicates work correctly.
Because these proofs concern non-ground HF formulas, theorem \isa{Sigma\_fm\_imp\_thm} does not help.
Required is an HF formalisation of operations on sequences, such as concatenation.
That in turn requires formalising further operations such as addition and set union.
These proofs (which are conducted largely in the HF calculus) total over 2,800 lines.
This total includes a library of results for truncating and concatenating sequences.
Here is a selection of the results proved.

Substitution preserves the value \isa{Zero}:

\begin{isabelle}
\isacommand{theorem}\ SubstTermP\_Zero:\ "\isacharbraceleft TermP\ t\isacharbraceright \ \isasymturnstile \ SubstTermP\ \isasymlceil Var\ v\isasymrceil \ t\ Zero\ Zero"
\end{isabelle}
On terms constructed using \isa{Eats} 
(recall that \isa{Q\_Eats} constructs the \emph{code} of an \isa{Eats} term),
substitution performs the natural recursion.

\begin{isabelle}
\isacommand{theorem}\ SubstTermP\_Eats:\isanewline
\ \ "\isacharbraceleft SubstTermP\ v\ i\ t1\ u1,\ SubstTermP\ v\ i\ t2\ u2\isacharbraceright\isanewline
\ \ \ \isasymturnstile \ SubstTermP\ v\ i\ (Q\_Eats\ t1\ t2)\ (Q\_Eats\ u1\ u2)"
\end{isabelle}
This seemingly obvious result takes nearly 150 lines to prove. The sequences for both substitution computations are combined to form a new sequence, which must be extended to yield the claimed result and shown to be properly constructed.

Substitution preserves constants. This fact is proved by induction on the sequence buildup
of the constant~\isa{c}, using the previous two facts about \isa{SubstTermP}\@.

\begin{isabelle}
\isacommand{theorem}\ SubstTermP\_Const:\ "\isacharbraceleft ConstP\ c,\ TermP\ t\isacharbraceright \ \isasymturnstile \ SubstTermP\ \isasymlceil Var\ w\isasymrceil \ t\ c\ c"
\end{isabelle}

Each recursive case of a syntactic predicate must be verified using the techniques
outlined above for \isa{SubstTermP\_Eats}. Even when there is only a single operand, 
as in the following case for negation, the proof is around 100 lines. 
\begin{isabelle}
\isacommand{theorem}\ SubstFormP\_Neg:\ "\isacharbraceleft SubstFormP\ v\ i\ x\ y\isacharbraceright \ \isasymturnstile \ SubstFormP\ v\ i\ (Q\_Neg\ x)\ (Q\_Neg\ y)"
\end{isabelle}
A complication is that
\isa{LstSeqP} accepts sequences that are longer than necessary, and these must be truncated to a given length before they can be extended. These lengthy arguments must be repeated for each similar proof. So, for the third time, one of our chief tools is cut and paste. 

Exactly the same sort of reasoning can be used to show that proofs can be combined as expected in order to apply inference rules. The following theorem expresses the Hilbert-Bernays
derivability condition~(D2):
\begin{isabelle}
\isacommand{theorem}\ PfP\_implies\_ModPon\_PfP:\ "\isasymlbrakk H\ \isasymturnstile \ PfP\ (Q\_Imp\ x\ y);\ H\ \isasymturnstile \ PfP\ x\isasymrbrakk \ \isasymLongrightarrow \ H\ \isasymturnstile \ PfP\ y"
\end{isabelle}
Now only one task remains: to show condition (D3).

\subsection{Making Sense of Quoted Values}

As mentioned in Sect.\ts\ref{sec:G-I}, making sense of expressions like 
$x=y\to\Pf{\quot{\boldsymbol{x}=\boldsymbol{y}}}$ requires a function $Q$ such that $Q(x)=\quot{x}$:
\begin{align*}
Q(0) & = \quot{0} = 0\\
Q(x\lhd y) &= \tuple{\quot{\lhd},Q(x),Q(y)}
\end{align*}
Trying to make this definition unambiguous, \'Swierczkowski \cite{swierczkowski-finite} sketches a total ordering on sets, but the technical details are complicated and incomplete. The same ordering can be defined via the function $f:\text{HF}\to \mathbb{N}$ such that $f(x)=\sum\,\{2^{f(y)}\mid y\in x\}$. It is intuitively clear, but formalising the required theory in HF would be laborious. It turns out that we do not need a canonical term $\boldsymbol{x}$ or a function~$Q$\@. We only need a relation: \isa{QuoteP} relates a set~$x$ to (the codes of) the terms that denote~$x$.  

The relation \isa{QuoteP} can be defined using precisely the same methods as we have seen above
for recursive functions, via a sequence buildup. The following facts can be proved using
the methods described in the previous sections.
\begin{isabelle}
\isacommand{lemma}\ QuoteP\_Zero:\ "\isacharbraceleft \isacharbraceright \ \isasymturnstile \ QuoteP\ Zero\ Zero"\vskip1ex
\isacommand{lemma}\ QuoteP\_Eats:\isanewline
\ \ "\isacharbraceleft QuoteP\ t1\ u1,\ QuoteP\ t2\ u2\isacharbraceright \ \isasymturnstile \ QuoteP\ (Eats\ t1\ t2)\ (Q\_Eats\ u1\ u2)"
\end{isabelle}
It is also necessary to prove (by induction within the HF calculus) 
that for every~$x$ there exists some term $\boldsymbol{x}$.
\begin{isabelle}
\isacommand{lemma}\ exists\_QuoteP:\isanewline
\ \ \isakeyword{assumes}\ j:\ "atom\ j\ \isasymsharp \ x"\ \ \isakeyword{shows}\ "\isacharbraceleft \isacharbraceright \ \isasymturnstile \ Ex\ j\ (QuoteP\ x\ (Var\ j))"
\end{isabelle}
We need similar results for all of the predicates involved in concatenating two sequences. They essentially prove that the corresponding pseudo-functions are total.

Now we need to start connecting these results with those of the previous section, 
which (following \'Swierczkowski) are proved for constants in general, although they are 
needed only for the outputs of \isa{QuoteP}\@. An induction in HF on the sequence buildup
proves that these outputs satisfy \isa{ConstP}\@.

\begin{isabelle}
\isacommand{lemma}\ QuoteP\_imp\_ConstP:\ "\isacharbraceleft \ QuoteP\ x\ y\ \isacharbraceright \ \isasymturnstile \ ConstP\ y"
\end{isabelle}
This is obvious, because \isa{QuoteP} involves only \isa{Zero} and \isa{Q\_Eats}, which construct quoted sets. Unfortunately, the proof requires the usual reasoning about sequences in order to show basic facts about \isa{ConstP}, which takes hundreds of lines. 

The main theorem from the previous section included the set of formulas
\begin{isabelle}
\ \ \ \ \isacharbraceleft ConstP\ (F\ i)\ |\ i.\ i\ \isasymin \ V\isacharbraceright
\end{isabelle}
on the left of the turnstile, representing the assumption that all introduced variables denoted constants. Now we can replace this assumption by one expressing that the relation \isa{QuoteP} holds between each pair of old and new variables.

\begin{isabelle}
\isacommand{definition}\ quote\_all\ ::\ "[perm,\ name\ set]\ \isasymRightarrow \ fm\ set"\isanewline
\ \ \isakeyword{where}\ "quote\_all\ p\ V\ =\ \isacharbraceleft QuoteP\ (Var\ i)\ (Var\ (p\ \isasymbullet \ i))\ |\ i.\ i\ \isasymin \ V\isacharbraceright 
\end{isabelle}

The relation \isa{QuoteP\ (Var\ i)\ (Var\ (p\ \isasymbullet \ i)} holds between
the variable \isa{i} and the renamed variable \isa{p\ \isasymbullet \ i}, for all \isa{i\ \isasymin \ V}\@.
Recall that \isa{p} is a permutation on variable names.
By virtue of the theorem \isa{QuoteP\_imp\_ConstP}, 
we obtain a key result, which will be used heavily in subsequent proofs for reasoning about coded formulas and the Pf predicate.
\begin{isabelle}
\isacommand{theorem}\ quote\_all\_PfP\_ssubst:\isanewline
\ \ \isakeyword{assumes}\ \isasymbeta:\ "\isacharbraceleft \isacharbraceright \ \isasymturnstile \ \isasymbeta "\isanewline
\ \ \ \ \ \ \ \isakeyword{and}\ V:\ "V\ \isasymsubseteq \ Vs"\isanewline
\ \ \ \ \ \ \ \isakeyword{and}\ s:\ "supp\ \isasymbeta \ \isasymsubseteq \ atom\ `\ Vs"\isanewline
\ \ \ \ \isakeyword{shows}\ \ \ \ "quote\_all\ p\ V\ \isasymturnstile \ PfP\ (ssubst\ \isasymlfloor \isasymbeta \isasymrfloor V\ V\ F)"
\end{isabelle}
In English: Let \isa{\isasymturnstile \ \isasymbeta} be a theorem of HF whose free variables belong to the set \isa{Vs}.
Take the code of this theorem, \isa{\isasymlfloor \isasymbeta \isasymrfloor},
and replace some subset \isa{V\ \isasymsubseteq \ Vs} of its free variables  by new variables
constrained by the \isa{QuoteP} relation.
The result, \isa{ssubst\ \isasymlfloor \isasymbeta \isasymrfloor V\ V\ F}, 
satisfies the provability predicate.

The reader of even a very careful presentation of G\"odel's second incompleteness theorem, such as Grandy \cite{grandy-advanced}, will look in vain for a clear and rigorous treatment of the $\boldsymbol{x}$ (or $\underline{x}$) convention. Boolos \cite{boolos-provability} comes very close with his Bew$[F]$ notation,
but he is quite wrong to state ``notice that Bew$[F]$ has the same variables free as [the formula] $F$'' 
\cite[p.\ts45]{boolos-provability} when in fact they have no variables in common. Even \'Swierczkowski's highly detailed account is at best ambiguous.
He consistently uses function notation, but his carefully-stated guidelines for replacing occurrences of pseudo-functions by quantified formulas \cite[Sect.\ts5]{swierczkowski-finite} are not relevant here. (This problem only became clear after a time-consuming attempt at a formalisation on that basis.) My companion paper \cite{paulson-incompl-logic}, which is aimed at logicians, includes a more detailed discussion of these points. 
It concludes that these various notations obscure not only the formal details of the proof but also the very intuitions they are intended to highlight.

\subsection{Proving $\vdash \alpha \to \Pf(\quot{\alpha})$}
 \label{sec:proving-reification}

We now have everything necessary to prove condition (D3):
\[ \text{If $\alpha$ is a strict $\Sigma$ sentence, then } \vdash \alpha \to \Pf(\quot{\alpha}) \]

The proof will be by induction on the structure of $\alpha$.
As stated in Sect.\ts\ref{sec:sigma} above, a strict $\Sigma$ formula
has the form $x\in y$, $\alpha\lor\beta$, $\alpha\land\beta$,  
$\exists x\,\alpha$ or $(\forall x\in y)\,\alpha$. 
Therefore, the induction will include one single base case,
\begin{align}
x\in y\to\Pf{\quot{\boldsymbol{x}\in \boldsymbol{y}}},
\label{eqn:base}
\end{align}
along with inductive steps for disjunction, conjunction, etc. 

\subsubsection{The Propositional Cases}

The propositional cases are not difficult, but are worth examining as a warmup exercise. From the induction hypotheses
$\vdash \alpha \to \Pf(\quot{\alpha})$ and 
$\vdash \beta  \to \Pf(\quot{\beta})$,
we must show
\begin{align*}
\vdash \alpha\lor\beta &\to \Pf(\quot{\alpha\lor\beta}) \text{ and}\\
\vdash \alpha\land\beta  &\to \Pf(\quot{\alpha\land\beta}). 
\end{align*}
Both of these cases are trivial by propositional logic, given the lemmas $\vdash \Pf(\quot{\alpha}) \to \Pf(\quot{\alpha\lor\beta})$, $\vdash \Pf(\quot{\beta}) \to \Pf(\quot{\alpha\lor\beta})$ and
\begin{align}
 \vdash \Pf(\quot{\alpha}) \to \Pf(\quot{\beta})  &\to \Pf(\quot{\alpha\land\beta})  \label{eqn:land}
\end{align} 
Proving (\ref{eqn:land}) directly from the definitions would need colossal efforts, but there is a quick proof. The automation of the HF calculus is good enough to prove tautologies, and from $\vdash \alpha \to \beta \to \alpha\land\beta$, the proof formalisation condition%
\footnote{Of Sect.\ts\ref{sec:deduction}, but via the substitution theorem \isa{quote\_all\_PfP\_ssubst} proved above. The induction concerns generalised formulas involving pseudo-coding: \isa{PfP\ (ssubst\ \isasymlfloor \isasymalpha \isasymrfloor V\ V\ F)}.}
yields
\[ \vdash \Pf(\quot{\alpha \to \beta \to \alpha\land\beta})  \]
Finally, the Hilbert-Bernays derivability condition~(D2) yields the desired lemma, (\ref{eqn:land}). This trick is needed whenever we need to do propositional reasoning with $\Pf$.

\subsubsection{The Equality and Membership Cases}

Now comes one of the most critical parts of the formalisation. Many published proofs \cite{boolos-provability,swierczkowski-finite} of the second completeness theorem use the following lemma:
\begin{gather}
x=y\to\Pf{\quot{\boldsymbol{x}=\boldsymbol{y}}} \label{eqn:eqstar}
\end{gather}
This in turn is proved using a lemma stating that $x=y$ implies $\boldsymbol{x}=\boldsymbol{y}$,
which is false here: we have defined \isa{QuoteP} only as a relation, and even $\{0,1\}$ can be written in two different ways. 
Nevertheless, the statement~(\ref{eqn:eqstar})
is clearly true: if $\boldsymbol{x}$ and $\boldsymbol{y}$ are constant terms
denoting $x$ and $y$, respectively, where $x=y$, then 
$\Pf\quot{\boldsymbol{x}=\boldsymbol{y}}$ holds. The equivalence of two different ways of writing a finite set should obviously be provable. This problem recalls the situation described in \ref{sec:sigma} above, and the induction used to prove \isa{Subset\_Mem\_sf\_lemma}. The solution, once again, is to prove the conjunction
\[ (x\in y\to\Pf{\quot{\boldsymbol{x}\in \boldsymbol{y}}}) \land
   (x\subseteq y\to\Pf{\quot{\boldsymbol{x}\subseteq \boldsymbol{y}}}) \]
by induction (in HF) on the sum of the sizes of $\boldsymbol{x}$ and $\boldsymbol{y}$.
The proof is huge (nearly 340 lines), with eight universal quantifiers in the induction formula,
each of which must be individually instantiated in order to apply an induction hypothesis.
\begin{isabelle}
\ \ \ \ \ \ \ \isasymturnstile \ All\ i\ (All\ i'\ (All\ si\ (All\ li\ (All\ j\ (All\ j'\ (All\ sj\ (All\ lj\isanewline
\ \ \ \ \ \ \ \ \ \ (SeqQuoteP\ (Var\ i)\ (Var\ i')\ (Var\ si)\ (Var\ li)\ IMP\isanewline
\ \ \ \ \ \ \ \ \ \ \ SeqQuoteP\ (Var\ j)\ (Var\ j')\ (Var\ sj)\ (Var\ lj)\ IMP\isanewline
\ \ \ \ \ \ \ \ \ \ \ HaddP\ (Var\ li)\ (Var\ lj)\ (Var\ k)\ IMP\isanewline
\ \ \ \ \ \ \ \ \ \ \ \ \ (\ (Var\ i\ IN\ Var\ j\ IMP\ PfP\ (Q\_Mem\ (Var\ i')\ (Var\ j')))\ \ AND\isanewline
\ \ \ \ \ \ \ \ \ \ \ \ \ \ \ (Var\ i\ SUBS\ Var\ j\ IMP\ PfP\ (Q\_Subset\ (Var\ i')\ (Var\ j'))))))))))))"
\end{isabelle}
Using \isa{SeqQuoteP} (which describes the sequence computation of \isa{QuoteP}) gives access to a size measure for the two terms, which are here designated \isa{i} and~\isa{j}. Their sizes, \isa{li} and~\isa{lj}, are added using \isa{HaddP}, which is simply addition as defined in the HF calculus. (This formalisation of addition is also needed for reasoning about sequences.) Their sum, \isa{k}, is used as the induction variable.

Although the second half of the conjunction suffices to prove (\ref{eqn:eqstar}),
it is never needed outside of the induction, and neither is (\ref{eqn:eqstar}) itself. All we need is (\ref{eqn:base}).
And so we reach the next milestone.

\begin{isabelle}
\isacommand{lemma}\ \isanewline
\ \ \isakeyword{assumes}\ "atom\ i\ \isasymsharp \ (j,j',i')"\ "atom\ i'\ \isasymsharp \ (j,j')"\ "atom\ j\ \isasymsharp \ (j')"\ \isanewline
\ \ \isakeyword{shows}\ QuoteP\_Mem\_imp\_QMem:\isanewline
\ \ \ \ \ \ \ \ "\isacharbraceleft QuoteP\ (Var\ i)\ (Var\ i'),\ QuoteP\ (Var\ j)\ (Var\ j'),\ Var\ i\ IN\ Var\ j\isacharbraceright \isanewline
\ \ \ \ \ \ \ \ \ \isasymturnstile \ PfP\ (Q\_Mem\ (Var\ i')\ (Var\ j'))"\isanewline
\ \ \ \ \isakeyword{and}\ QuoteP\_Mem\_imp\_QSubset:\isanewline
\ \ \ \ \ \ \ \ "\isacharbraceleft QuoteP\ (Var\ i)\ (Var\ i'),\ QuoteP\ (Var\ j)\ (Var\ j'),\ Var\ i\ SUBS\ Var\ j\isacharbraceright \isanewline
\ \ \ \ \ \ \ \ \ \isasymturnstile \ PfP\ (Q\_Subset\ (Var\ i')\ (Var\ j'))"
\end{isabelle}

Turning to the main induction on $\alpha$, the notoriously ``delicate'' \cite[p.\ts48]{boolos-provability}
case is bounded universal quantification. Many of the delicate points here are connected with
the way the nominal approach is used. We need to maintain and extend a permutation
relating old and new variable names. Such complexities are evident in mathematical texts,
in their treatment of variable names \cite[Lemma~9.7]{swierczkowski-finite}.

\begin{isabelle}
\isacommand{lemma}\ (\isakeyword{in}\ quote\_perm)\ quote\_all\_Mem\_imp\_All2:\isanewline
\ \ \isakeyword{assumes}\ IH:\ "insert\ (QuoteP\ (Var\ i)\ (Var\ i'))\ (quote\_all\ p\ Vs)\ \isanewline
\ \ \ \ \ \ \ \ \ \ \ \ \ \ \ \isasymturnstile \ \isasymalpha \ IMP\ PfP\ (ssubst\ \isasymlfloor \isasymalpha \isasymrfloor (insert\ i\ Vs)\ (insert\ i\ Vs)\ Fi)"\isanewline
\ \ \ \ \ \ \isakeyword{and}\ "supp\ (All2\ i\ (Var\ j)\ \isasymalpha )\ \isasymsubseteq \ atom\ `\ Vs"\ \isanewline
\ \ \ \ \ \ \isakeyword{and}\ j:\ "atom\ j\ \isasymsharp \ (i,\isasymalpha )"\ \isakeyword{and}\ i:\ "atom\ i\ \isasymsharp \ p"\ \isakeyword{and}\ i':\ "atom\ i'\ \isasymsharp \ (i,p,\isasymalpha )"\ \isanewline
\ \ \isakeyword{shows}\ "insert\ (All2\ i\ (Var\ j)\ \isasymalpha )\ (quote\_all\ p\ Vs)\isanewline
\ \ \ \ \ \ \ \ \ \ \isasymturnstile \ PfP\ (ssubst\ \isasymlfloor All2\ i\ (Var\ j)\ \isasymalpha \isasymrfloor Vs\ Vs\ F)"
\end{isabelle}

The final case, for the existential quantifier, is also somewhat complicated to formalise. The details are again mostly connected with managing free and bound variable names using nominal methods, and are therefore omitted. We can conclude our discussion of the inductive argument by viewing the final result:
\begin{isabelle}
\isacommand{lemma}\ star:\isanewline
\ \ \isakeyword{assumes}\ "ss\_fm\ \isasymalpha "\ \ "finite\ V"\ \ "supp\ \isasymalpha \ \isasymsubseteq \ atom\ `\ V"\isanewline
\ \ \ \ \isakeyword{shows}\ "insert\ \isasymalpha \ (quote\_all\ p\ V)\ \isasymturnstile \ PfP\ (ssubst\ \isasymlfloor \isasymalpha \isasymrfloor V\ V\ F)"
\end{isabelle}

Condition (D3) now follows easily, since the formula $\alpha$ is then a sentence.
Although some technical conditions (involving variable names and permutations) have been omitted
from the previous two theorems, our main result below appears exactly as it was proved.
Of course, $\alpha\vdash\Pf\quot\alpha$ is equivalent to ${}\vdash\alpha\to\Pf\quot\alpha$.

\begin{isabelle}
\isacommand{theorem}\ Provability:\ \isanewline
\ \ \isakeyword{assumes}\ "Sigma\_fm\ \isasymalpha "\ "ground\_fm\ \isasymalpha "\ \isanewline
\ \ \ \ \isakeyword{shows}\ "\isacharbraceleft \isasymalpha \isacharbraceright \ \isasymturnstile \ PfP\ \isasymlceil \isasymalpha \isasymrceil "
\end{isabelle}

\section{G\"odel's Second Incompleteness Theorem} \label{sec:G-II}

The final steps of the second incompleteness theorem can be seen in Fig.\ts\ref{fig:G-II}.
The diagonal formula, $\delta$, is obtained via the first incompleteness theorem.
Then we can quickly establish both $\Pf\quot\delta\vdash\Pf(\quot{\Pf\quot\delta})$ 
and $\Pf\quot\delta\vdash\Pf(\quot{\neg\Pf\quot\delta})$.
These together imply $\Pf\quot\delta\vdash\Pf(\quot{\bot})$ using a variant of condition~(D2).
It follows that if the system proves its own consistency, then it also proves 
${}\vdash\neg \Pf\quot\delta$ and therefore ${}\vdash\delta$,
a contradiction.

\'Swierczkowski \cite{swierczkowski-finite} presents a few other results which have not
been formalised here. These include a refinement of the incompleteness theorem (credited to Reinhardt) and a theory of a linear order on the HF sets, but recall that  claim~(\ref{eqn:eqstar}) can be proved without using any such ordering. The approach adopted here undoubtedly involves less effort than formalising the ordering in the HF calculus.

The total proof length of nearly 12,400 lines comprises around 4,600 lines for the second theorem
and 7,700 lines for the first.%
\footnote{Prior to polishing and removing unused material, the proof totalled 17,000 lines.}
(One could also include 2,700 lines for HF set theory itself, 
but we would not count the standard libraries of natural numbers if they were used as the
basis of the proof.)
O'Connor's proof comprises 47,000 lines of Coq, while Shankar's takes 20,000 lines \cite[p.\ts139]{shankar94} and Harrison's \cite{harrison-goedel-length} is a miniscule 4,400 lines of HOL Light. Recall that none of these other proofs include the second incompleteness theorem.

But comparisons are almost meaningless because of the enormous differences among the formalisations.
Shankar wrote (and proved to be representable) a LISP interpreter for coding up the metatheory \cite{shankar94}; this was a huge effort, but then the various coding functions could then be written in LISP without further justification. He also used HF for coding, presumably because of its similarity to LISP S-expressions.
O'Connor formalised a very general syntax for first-order logic \cite{oconnor-incompleteness}. He introduced a general inductive definition of the primitive 
recursive functions, but proving specific functions to be primitive recursive turned out to be extremely difficult \cite[Sect.\ts5.3]{oconnor-phd}. Harrison has not published a paper describing his formalisation, but devotes a few pages to G\"odel's theorems in his \textit{Handbook of Practical Logic} \cite[p.\ts546--555]{harrison-handbook}, including extracts of HOL Light proofs. He defines $\Sigma_1$ and $\Pi_1$ formulas and quotes some meta-theoretical results relating truth and provability. 

\begin{figure}
\begin{isabelle}
\isacommand{theorem}\ Goedel\_II:\isanewline
\ \ \isakeyword{assumes}\ "\isasymnot \ \isacharbraceleft \isacharbraceright \ \isasymturnstile \ Fls"\isanewline
\ \ \ \ \isakeyword{shows}\ "\isasymnot \ \isacharbraceleft \isacharbraceright \ \isasymturnstile \ Neg\ (PfP\ \isasymlceil Fls\isasymrceil )"\isanewline
\isacommand{proof}\ -\isanewline
\ \ \isacommand{from}\ assms\ Goedel\_I\ \isacommand{obtain}\ \isasymdelta \ \isanewline
\ \ \ \ \isakeyword{where}\ diag:\ "\isacharbraceleft \isacharbraceright \ \isasymturnstile \ \isasymdelta \ IFF\ Neg\ (PfP\ \isasymlceil \isasymdelta \isasymrceil )"\ \ "\isasymnot \ \isacharbraceleft \isacharbraceright \ \isasymturnstile \ \isasymdelta "\isanewline
\ \ \ \ \ \ \isakeyword{and}\ gnd:\ \ "ground\_fm\ \isasymdelta "\isanewline
\ \ \ \ \isacommand{by}\ metis\isanewline
\ \ \isacommand{have}\ "\isacharbraceleft PfP\ \isasymlceil \isasymdelta \isasymrceil \isacharbraceright \ \isasymturnstile \ PfP\ \isasymlceil PfP\ \isasymlceil \isasymdelta \isasymrceil \isasymrceil "\isanewline
\ \ \ \ \isacommand{by}\ (auto\ simp:\ Provability\ ground\_fm\_aux\_def\ supp\_conv\_fresh)\isanewline
\ \ \isacommand{moreover}\ \isacommand{have}\ "\isacharbraceleft PfP\ \isasymlceil \isasymdelta \isasymrceil \isacharbraceright \ \isasymturnstile \ PfP\ \isasymlceil Neg\ (PfP\ \isasymlceil \isasymdelta \isasymrceil )\isasymrceil "\isanewline
\ \ \ \ \isacommand{apply}\ (rule\ MonPon\_PfP\_implies\_PfP\ [OF\ \_\ gnd])\isanewline
\ \ \ \ \isacommand{apply}\ (metis\ Conj\_E2\ Iff\_def\ Iff\_sym\ diag(1))\isanewline
\ \ \ \ \isacommand{apply}\ (auto\ simp:\ ground\_fm\_aux\_def\ supp\_conv\_fresh)\ \isanewline
\ \ \ \ \isacommand{done}\isanewline
\ \ \isacommand{moreover}\ \isacommand{have}\ "ground\_fm\ (PfP\ \isasymlceil \isasymdelta \isasymrceil )"\isanewline
\ \ \ \ \isacommand{by}\ (auto\ simp:\ ground\_fm\_aux\_def\ supp\_conv\_fresh)\isanewline
\ \ \isacommand{ultimately}\ \isacommand{have}\ "\isacharbraceleft PfP\ \isasymlceil \isasymdelta \isasymrceil \isacharbraceright \ \isasymturnstile \ PfP\ \isasymlceil Fls\isasymrceil "\ \isacommand{using}\ PfP\_quot\_contra\ \ \isanewline
\ \ \ \ \isacommand{by}\ (metis\ (no\_types)\ anti\_deduction\ cut2)\isanewline
\ \ \isacommand{thus}\ "\isasymnot \ \isacharbraceleft \isacharbraceright \ \isasymturnstile \ Neg\ (PfP\ \isasymlceil Fls\isasymrceil )"\isanewline
\ \ \ \ \isacommand{by}\ (metis\ Iff\_MP2\_same\ Neg\_mono\ cut1\ diag)\isanewline
\isacommand{qed}%
\end{isabelle}
\caption{G\"odel's Second Incompleteness Theorem}
\label{fig:G-II}
\end{figure}

This project took approximately one year, in time left available after fulfilling 
teaching and administrative duties. The underlying set theory took only two weeks to formalise.
The G\"odel development up to the proof formalisation condition took another five months.
From there to the first incompleteness theorem took a further two months, 
mostly devoted to proving single valued properties.
Then the second incompleteness theorem took a further four months, 
including much time wasted due to misunderstanding this perplexing material.
Some material has since been consolidated or streamlined. The final version is available online \cite{Incompleteness-AFP}.

\subsection{The Lengths of Proofs}

\begin{figure}[htbp]
   \centering
   \includegraphics[scale=0.6]{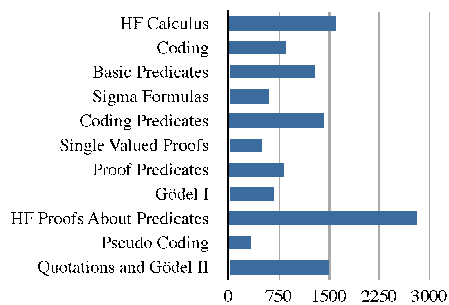} 
   \caption{Sizes of Constituent Theories}
   \label{fig:sizes}
\end{figure}

Figure~\ref{fig:sizes} depicts the sizes of the Isabelle/HOL theories making up various sections of the proof development. The first theorem takes up the bulk of the effort. Apart from the massive HF proofs about predicates, which are mostly of obvious properties, the second theorem appears to be a fairly easy step given the first. Why then has it not been formalised until now?
A reasonable guess is that previous researchers were not aware of 
\'Swierczkowski's \cite{swierczkowski-finite} elaborate development.
The most detailed of the previous proofs \cite{boolos-provability,grandy-advanced} 
left too much to the imagination, and even \'Swierczkowski makes some errors. He devotes much of his Sect.\ts7 to  proofs concerning substitution for (non-existent) pseudo-terms analogous to~$\boldsymbol{x}$. Recall that pseudo-terms are merely a notational shorthand to allow function syntax, and are not actual terms. Finally, there was the critical issue of the ordering on the HF sets. Solving these mysteries required much thought, and the first completed formalisation included thousands of lines of proofs belonging to aborted attempts.

A discussion of the de Bruijn coefficient (the expansion in size entailed by the process of formalisation) is always interesting, but difficult to do rigorously. In this case, the formalisation required proving a great many theorems that were not even hinted at in the source text, for example, setting up a usable sequent calculus (\'Swierczkowski merely gives the rudiments of a Hilbert system), or proving that the various codings of syntax actually work (virtually all authors take this for granted), or proving that ``functions'' are functions. The hundreds of lines of single-step HF calculus proofs are the single largest contributor to the size of this development, and such things never appear in standard presentations of the incompleteness theorems.

A crude calculation yields 30 pages at 35 lines per page or 1050 lines for \'Swier\-czkowski's proof, compared with 12,400 lines of Isabelle, for a de Bruijn factor of 12. Focusing on just the proof of the first incompleteness theorem (after the preliminary developments), we have about 80 lines of informal text and 671 lines of Isabelle, giving a factor of 8.4; that includes 150 lines in the Isabelle script to prove that functions are single valued. 

A further issue is the heavy use of cut and paste in the HF calculus proofs. Better automation for HF could help, but spending time to develop a tactic that will only be called a few times is hard to justify. An alternative idea is to define higher-order operators for the sequence constructions, which could be proved to be functional once and for all. However, higher-order operators are difficult to define using nominal syntax. Perhaps it could be attempted using naive variables.

\subsection{The Formalisation of Variable Binding}

The role of bound variable syntax remains unclear. Shankar \cite{shankar94,shankar-deBruijn}
used de Bruijn variables to formalise the Church-Rosser theorem but not the incompleteness theorem.
Harrison did not use them either.
O'Connor also used traditional syntactic bound variables, but complained 
about huge complications concerning substitution (recall Sect.\ts\ref{sec:nominal}).
The present development uses the nominal package, easing many proofs, but at a price:
over 900 lines involve freshness specifications, and around 70 lemmas involving freshness are proved.
Moreover, many proof steps are slow. While the project was underway, proof times
taking minutes were not unusual. Even after recent improvements to the nominal package,
they can take tens of seconds. Additional performance improvements would be welcome, 
as well as a more concise notation for freshness conditions when many new names are needed.

In fairness to the nominal approach, explicit variable names would also have to be fresh and analogous assertions would be necessary, along with some procedure for calculating new names numerically and proving them to be fresh. The effort may be similar either way, but the nominal approach is more abstract and natural: who after all refers to specific, calculated variable names in textbook proofs? 

My first attempt to formalise the incompleteness theorems used explicit names. Then, substitution on formulas was only available as a relation, and many proofs required numerical operations on variable names. This effort would have multiplied considerably for the second incompleteness theorem. Using de Bruijn indices for HF syntax was not attractive: I had previously formalised G\"odel's definition of the constructible sets and his proof of the relative consistency of the axiom of choice \cite{paulson-consistency}. Here it was also necessary to define a great many predicates within an encoding of first-order logic. This work was done in Isabelle/ZF, a version of Isabelle for axiomatic set theory. I used de Bruijn indices in these definitions, but the loss of readability was a severe impediment to progress. 

It is worth investigating how this formalisation would be affected by the change to
another treatment of variable binding. As regards the G\"odel numbering of formulas,
the use of de Bruijn variables can be called an unqualified success. It was easy to set up and all necessary properties were proved without great difficulties.

\subsection{On Verifying Proof Assistants}

In a paper entitled ``Towards Self-verification of HOL Light'', Harrison says,
\begin{quote}
G\"odel's second incompleteness theorem tells us that [a logical system] 
cannot prove its own consistency in any way at all \ldots. So, regardless of implementation details, if we want to prove the consistency of a proof checker, we need to use a logic that in at least some respects goes beyond the logic the checker itself supports.  \cite[p.\ts179]{harrison-self}
\end{quote}
This statement is potentially misleading, and has given rise to the mistaken view that it is impossible to verify a proof checker in its own logic.

Harrison's aim is to prove that HOL Light cannot prove the theorem \texttt{FALSE}, and this indeed requires proving the consistency of higher-order logic itself. 
Unfortunately, most consistency proofs are unsatisfactory because they more or less assume the desired conclusion: they are thinly disguised versions of the tautology $\mathop{\rm Con}(L)\land L\to \mathop{\rm Con}(L)$. 
This is a consequence of the second incompleteness theorem, 
since the consistency of~$L$ can only be proved in a strictly stronger formal system. 

Mathematicians accept strong formal systems, such as ZF set theory, with little justification other than intuition and experience. 
Moreover, they examine very strong further axioms. 
The axiom of constructibility is an instructive case: it is known to be relatively consistent with respect to the axioms of set theory,
but it is not generally accepted as true \cite[p.\ts170]{kunen80}.
The standard ZF axioms are generally regarded as true, although they cannot even be proved to be consistent.
Thus we have no good way of proving consistency, and yet consistency does not guarantee truth.

This situation calls for a separation of concerns. The builders of verification tools should be concerned with the correctness of their code, but the correctness of the underlying formal calculus is the concern of logicians.
Harrison notes that ``almost all implementation bugs in HOL Light and other versions of HOL have involved variable renaming'' \cite[p.\ts179]{harrison-self}, and this type of issue should be our focus. 
Verifying a proof assistant involves verifying that it implements a data structure for the assertions of the formal calculus and that it satisfies a commuting diagram relating deductions on the implemented assertions with the corresponding deductions in the calculus. 
G\"odel's theorems have no relevance here.

\section{Conclusions} \label{sec:conclusions}

The main finding is simply that G\"odel's second incompleteness theorem can be proved with a relatively modest effort, in only a few months starting with a proof of the first incompleteness theorem. While the nominal approach to syntax is clearly not indispensable, it copes convincingly with a development of this size and complexity. The use of HF set theory as an alternative to Peano arithmetic is clearly justified, eliminating the need to formalise basic number theory within the embedded calculus; the necessary effort to do that would greatly exceed the difficulties (mentioned in Sect.\ts\ref{sec:proving-reification} above) caused by the lack of a simple canonical ordering on HF sets.

Many published proofs of the incompleteness theorems replace technical proofs by vague appeals to Church's thesis. Boolos \cite{boolos-provability} presents a more detailed and careful exposition, but still leaves substantial gaps. Even the source text \cite{swierczkowski-finite} for this project, although written with great care, had problems: a significant gap (concerning the canonical ordering of HF sets), a few minor ones (concerning $\Sigma$ formulas, for example), and pages of material that are, at the very least, misleading. These remarks are not intended as criticism but as objective observations of the complexity of this material, with its codings of codings. A complete formal proof, written in a fairly readable notation, should greatly clarify the issues involved in these crucially important theorems.


\section*{Acknowledgment}
Jesse Alama drew my attention to \'Swierczkowski \cite{swierczkowski-finite}, the source material for this project. Christian Urban assisted with nominal aspects of some of the proofs, even writing code. Brian Huffman provided the core formalisation of type~\isa{hf}. Dana Scott offered advice and drew my attention to Kirby \cite{kirby-addition}. Matt Kaufmann and the referees made many insightful comments.

\bibliographystyle{plain}\raggedright\small

\bibliography{string,atp,funprog,general,isabelle,theory,crossref}
\end{document}